\documentclass[prd,aps,10pt,twocolumn,floats,floatfix,nofootinbib] {revtex4-1}

\usepackage{graphicx}
\usepackage{dcolumn}
\usepackage{bm}
\usepackage{graphics}
\usepackage{slashed}
\usepackage{amssymb}
\usepackage{natbib}
\usepackage{amsmath}
\usepackage{url}
\usepackage[usenames,dvipsnames,svgnames,table]{xcolor}
\usepackage{color}
\usepackage{xcolor}
\usepackage{tabularx}
\usepackage{hyperref}
\usepackage{enumitem}
\usepackage{soul}

\begin{document}
\title{Numerical simulations of divergence-type theories for conformal dissipative fluids}
\author{Pablo E. Montes${}^{1}$}\email{paloemontes@mi.unc.edu.ar}
\author{Marcelo E. Rubio$^{2\mbox{,}3}$}\email{mrubio@sissa.it}
\author{Oscar A. Reula${}^{1\mbox{,}4}$}\email{reula@famaf.unc.edu.ar}
\affiliation{\vspace{0.2cm}
${}^{1}$Instituto de F\'{\i}sica Enrique Gaviola (CONICET), Medina Allende y Haya de la Torre, s/n, Córdoba, Argentina\\
${}^{2}$SISSA, Via Bonomea 265, 34136 Trieste, Italy and INFN (Sezione di Trieste)\\
${}^{3}$IFPU - Institute for Fundamental Physics of the Universe, Via Beirut 2, 34014 Trieste, Italy
\\
${}^{4}$Facultad de Matem\'atica, Astronom\'ia, F\'isica y Computaci\'on, 
Ciudad Universitaria, (5000) C\'ordoba, Argentina}

\begin{abstract}
We report on the first numerical simulations of the symmetric--hyperbolic theory for conformal dissipative relativistic fluids developed in \cite{Rubio18}. In this theory, the information of the fluid dynamics is encoded in a scalar generating function which depends on three free parameters. By adapting the WENO-Z high-resolution shock-capturing central scheme, we present numerical solutions restricted to planar symmetry in Minkowski spacetime, from two qualitatively different initial data: a smooth bump and a discontinuous step. We perform a detailed exploration of the effect of the different parameters of the theory, and numerically assess the constitutive relations associated with the shear viscosity by analyzing the entropy production rate when shocks are produced.
\end{abstract}

\maketitle


\section{Introduction}
\label{sec-1}

The success of Relativistic Hydrodynamics as a framework for modeling matter and energy transport, has attracted the attention of the community since its origins. One of the main reasons for such a fame is perhaps its surprising versatility to be applied over a wide range of physical phenomena, at very different scales. In Astrophysics, for instance, it is used for modeling relativistic jets from the core of active galactic nuclei \cite{Netzer13,Blandford19}, micro-quasars \cite{Charlet21}, rotating black holes \cite{Meier03}, and gamma-ray burst central engines \cite{Peer15,Ramirez10}. It is also applied for describing the chemical and thermodynamical composition of the interior of compact stars \cite{Raposo19}, and even for exploring the dynamical structure of accretion disks around rotating black holes \cite{Stone01,Yuan14,Das07,Giustini19}. In particular, certain hydrodynamic models for the equation of state of the interior of compact objects turned out to be crucial in the new window that has been recently opened from the first direct detection of gravitational waves \cite{Abbott21,GW170817,Chatziioannou18,Landry20}.

Relativistic Hydrodynamics also models the micro-physics of strongly interacting matter produced in heavy-ion colliders \cite{Busza18,DEnterria06,McDonough20} like the LHC and the RHIC, making reliable predictions
about their most intimate structure \cite{Calzetta99,Peralta11,Calzetta14,Elias14}. Among the different experiments pursued in these accelerators is to probe new possible phases of nuclear matter at high energies. Assuming a Bjorken model for the collision, matter is expected to be a very low-mass plasma composed mainly of gluons and quarks. Both nuclei approach each other at ultra-relativistic speeds, producing a wake of very hot plasma which expands and cools down, eventually fragmenting into different particles. The latters are subjected to high energy scattering processes, which are subsequently detected in the laboratory. Although the thermal properties of the plasma may be described from perturbative QCD theory, in order to obtain a more quantitative description it is useful to describe it as a relativistic ideal fluid (applicable to matter in local thermodynamic equilibrium), provided a suitable equation of state. The reason for ``neglecting'' QCD theory at a first approximation is that hydrodynamic models offer a natural way to couple flow to pressure gradients in the transverse plane of the collision, reproducing experimental data in a surprising accurate way \cite{Ollitrault07,Jaiswal16}. Nevertheless, ideal hydrodynamics usually overestimates the anisotropy in the transverse flow. Including viscous effects, instead, yields a much better agreement, if the ratio between the fluid entropy and the shear viscosity keeps sufficiently small. One interesting example of this situation is the so-called \textit{elliptic flow}, which is characterized by the pressure anisotropy in the plane orthogonal to the flow velocity direction. Ideal hydrodynamics predicts a larger amount of elliptic flow compared to the one observed experimentally, suggesting that there should be some physical mechanism helping to ``isotropize’’ the system \cite{Calzetta199,Calzetta21}. This is indeed the role played by viscosity in this context, which motivates, among other arguments, the importance of having a ``well-behaved'' dissipative hydrodynamic theory.

The fundamental equations describing relativistic fluids are the local conservation of the energy-momentum tensor and the particle number density current \cite{geroch1990dissipative}. Such laws are the covariant generalization of the well-known energy and momentum conservation laws and the continuity equation for the mass in Newtonian fluid mechanics \cite{landau}. In order to close the system, an equation of state relating thermodynamic quantities such as the pressure, internal energy and energy density needs to be fixed. Although the choice of the equation of state could become enough sophisticated for taking into account chemical processes (such as molecular interactions, quantization, relativistic effects and/or nuclear processes), the most widely employed in astrophysical simulations are the ideal-gas, the polytropic and the radiation ones; the choice will always depend on the physics one is interested in modeling \cite{rezzolla2013relativistic}. 

In the context of General Relativity, the gravitational interaction of the fluid is obtained by coupling the hydrodynamic equations with the gravitational field (Einstein's) equations. However, in the majority of astrophysical scenarios (like accretion processes or propagation of relativistic jets) the ``test-fluid'' approximation is good enough to provide an accurate description of the underlying dynamics \cite{Font08,Font07,Carrasco17,Carrasco18Pulsar}, being the self-gravity of the fluid completely neglected in comparison to the background gravitational field, and the mass of the accreting fluid becomes usually much smaller than the rest mass of the compact object \cite{Komiss04,Fernandez18}.

The parabolic nature of the Navier-Stokes-Fourier system of equations for non-relativistic viscous fluids, implies that they cannot be naively extended to relativistic regimes. Motivated by this, a great deal of effort has been devoted into developing theories that allow an accurate description of dissipative fluids; i.e., theories which are (i) causal, (ii) stable, and (iii) whose initial-value problem is well-posed. This is still an open problem in the community, and the main reason is that there is a lot of freedom on how to model non-equilibrium dynamics when considering energy transport/dissipation effects \cite{Kovtun12}. The first attempts towards this huge goal date back to the works by Eckart \cite{Eckart40} and Landau-Lifshitz \cite{landau}, who proposed two particular ways to include dissipative corrections in the dynamical variables of the theory. However, these proposals did not come to fruition, because years later it was shown that they are not only unstable, but also ill-posed. This means that they admit modes which grow exponentially with the wavenumber and, moreover, such instabilities turn out to be generic, in the sense that they manifest with respect to any frame \cite{hiscock1983stability,Hiscock:1985zz}. 

Some years later, alternative ways for modeling dissipative effects on fluid systems came up, as the well-known Israel-Stewart \cite{Israel-Stew79} theory of extended thermodynamics. Although this formulation provided a stable and well-posed theory, other approaches were also developed. One of them was the so-called \textit{divergence-type} theories, originally introduced by Pennisi, Liu and Ruggieri \cite{liu1986relativistic}, including extra dynamical degrees of freedom from which they could describe dissipative processes, as energy transport through heat fluxes, entropy generation and viscosity effects. Later on, a deeper study of the well-posedness and stability of divergence-type theories was carried out by Geroch and Lindblom \cite{geroch1991causal,geroch1990dissipative}. In this approach, the dissipative effects are encoded in a new ``constitutive'' tensor field. The advantage of these theories is that, as a consequence of the symmetry of the energy-momentum tensor, all the information of the theory is contained in a single generating scalar function, whose different terms consider dissipative contributions to different orders. It is not difficult to see that considering first-order dissipative generating functions leads to theories that are only weakly-hyperbolic, and so have the pathologies shown by Hiscock and Lindblom \cite{Kreiss70}. Nevertheless, there exist some families of second-order generating functions leading to well-posed theories \cite{geroch1996partial}. By well-posedness we mean that, for any given initial data set in certain Banach space there exists a finite time $T$ such that: (i) during this time there exists a solution belonging to some other Banach space; (ii) that solution is unique; (iii) the solution is a continuous function of the initial data given (in the corresponding topologies where the data and solutions are defined) \cite{Hadamard1908}. Dynamical evolution is governed by the \textit{principal part} of the system of equations, which contains information about the propagation speeds of the different modes \cite{Kreiss70,geroch1996partial}. The modern way to elucidate this non-trivial mathematical condition is through the concept of \textit{hyperbolicity} \cite{Friedrichs54,Kreiss70,friedrichs1971systems,geroch1996partial}, that is, a set of algebraic conditions the principal part should satisfy for the system to be well-posed (or, as it is commonly known, strongly-hyperbolic).

An essential edge for the use of relativistic dissipative theories for solving concrete problems is the possibility of generating solutions through numerical simulations. Indeed, a large variety of numerical schemes for simulating hydrodynamic systems have been successfully built during the last decades \cite{Aloy99,Font00,Font02,Radice11,Radice12,Radice14}; mostly based on explicit finite difference upwind schemes; specifically designed to solve nonlinear hyperbolic systems of conservation laws \cite{Lax73} (and most of them originally developed from codes for solving non-relativistic hydrodynamics). These schemes implement approximate or exact Riemann solvers \cite{Roe97}, starting from the characteristic decomposition of the corresponding system of conservation equations and based on algorithms which are able to robustly capture sharp discontinuities along evolution (see \cite{Alcubierre08,rezzolla2013relativistic,leveque92,Lax73} detailed discussions, and references therein). Among the most popular numerical algorithms for evolving this kind of equations are the high-order and non-oscillatory central schemes \cite{Harten83,Blazek01,Versteeg07}. One of the nicest properties of central schemes is that they exploit the conservation form of the Lax-Wendroff \cite{LaxWen60} and Lax-Friedrichs schemes (see the book \cite{Thomas13} for a recent discussion), yielding the correct propagation speeds of all nonlinear waves appearing in the solution. This was reached without using Riemann solvers, resulting then in a high computational efficiency. After Lax's seminal work in the mid-50s \cite{Lax-num1-54}, there came up a huge variety of extensions aiming to enhance certain oscillatory behaviours near shocks or discontinuities, which were no proper of the physical situation of interest.

Very recently, a theory for relativistic fluids with first-order dissipative contributions was proposed by Bemfica, Disconzi, Noronha and Kovtun (BDNK) \cite{Bemfica21,Bemfica:2017wps,Kovtun12}. Unlike previous efforts, this theory does not present the generic instabilities reported by Hiscock and Lindblom for the Eckart and Landau-Lifshitz theories. After showing that the corresponding initial value problem is well-posed in Gevrey spaces (which is commonly known as the Leráy hyperbolicity \cite{Leray-Ohya67,Leray53}), results of stability, causality and strong hyperbolicity in Sobolev spaces \cite{friedrichs1954symmetric} where also reported. These results motivated a series of recent works \cite{Pretorius21,Pretorius22,Figueras22}, in which the BDNK theory was explored numerically, finding initial smooth configurations that develop shocks during evolution.

In this work, we present the first numerical solutions of the system of equations governing the dynamics of dissipative ultra-relativistic (or conformal) fluids, whose theory was previously developed in \cite{Rubio18}. This scheme constitutes, to the best of our knowledge, the first proposal for evolving second-order divergence-type dissipative fluid theories. To do so, we consider two sets of variables (which we refer as \textit{conservative} and \textit{fluid} variables), we invert the relation between them and reconstruct the corresponding fluxes in term of the conservative variables for the time evolution. After introducing the most general dynamical equations for second-order dissipative fluids, we consider the simplest dynamical case as a first exploration; i.e., a fluid propagating in only one spatial dimension and in flat space, by imposing rotational invariance in the plane perpendicular to it (sometimes referred as ``slab-symmetric'' configurations). These assumptions allow a considerable reduction of the degrees of freedom of the theory, and consequently a rather natural and intuitive first-step implementation towards the full 3D general case, over an arbitrary background geometry. 

This paper is organized as follows. Section \ref{sec-2} contains a brief review of the conformal theory developed in \cite{Rubio18}, introducing the fundamental concepts, identities and equations that will be used further. In Section \ref{sec-3}, the general (3+1)-decomposition is presented, and a discussion about the choice of suitable evolution variables is addressed. After particularizing to the 1+1 reduced system, and analyzing the corresponding characteristic structure, we introduce the numerical method we will use, as well as some details on the implementation. Section \ref{sec-4} is dedicated to the numerical results, code validation, convergence, and physical analysis of the solutions, comparing them with the ideal-fluid case. Finally, an overall discussion and final remarks are presented in Section \ref{sec-5}. 

Throughout this work, the signature convention for the spacetime metric is $(-,+,+,+)$. We use geometric units with $c = G = k_B = 1$, where $c$ is the speed of light in vacuum, $G$ is Newton's constant in four spacetime dimensions and $k_B$ is Boltzman's constant.


\section{Conformal Hydrodynamics in 4D}
\label{sec-2}


In this section, we review fundamental aspects of the family of conformally-invariant divergence-type fluid theories developed in \cite{Rubio18}, in order to fix notation and definitions which shall be further used\footnote{For details, we refer the reader to references \cite{pennisi87,geroch1990dissipative,geroch1991causal,geroch1995relativistic,geroch2001hyperbolic}.}. We also discuss the conformal weights of the different quantities involved in the theory, as well as the structure of the dynamical equations at thermodynamic equilibrium and their further implications.

\subsection{Preliminaries}

We consider fluid theories over a time-oriented 4--dimensional background spacetime $(\mathcal{M},g_{ab})$. The fluid degrees of freedom are encoded in two tensor fields, $T^{ab}$ and $A^{abc}$, which satisfy the conservation laws
\begin{eqnarray}
\nabla_a T^{ab} &=& 0, \label{Tabcons} \\
\nabla_a A^{abc} &=& I^{bc}. \label{dissipeq} 
\end{eqnarray}
The source $I^{bc}$ is a symmetric, trace-free, algebraic function of $T^{ab}$ and $A^{abc}$, and $\nabla_c$ is the covariant derivative compatible with $g_{ab}$. Equation (\ref{Tabcons}) states energy momentum conservation of the fluid, and (\ref{dissipeq}) governs the evolution of the dissipative degrees of freedom, providing also constitutive relations. 
Since we aim to give a general conformally invariant theory, no conservation of baryon density current will be considered along this work. By construction, $T^{ab}$ is symmetric, and $A^{abc}$ is symmetric and trace-free in the last two indices, namely $A^{abc}g_{bc} = 0$ and $A^{a[bc]} = 0$. We shall refer to $T^{ab}$ and $A^{abc}$ as the \textit{conservative} variables. 

It is also assumed the existence of an entropy density current, $S^a$, which is an algebraic function of the conservative variables, and as a consequence of (\ref{Tabcons})-(\ref{dissipeq}), it satisfies
\begin{equation}\label{entropycons}
\nabla_a S^a = \sigma \geq 0,
\end{equation}
where $\sigma$ is also algebraic in $T^{ab}$ and $A^{abc}$.

A remarkable consequence about the existence of an entropy current within this framework was pointed out earlier by Liu, Ruggieri and Pennisi \cite{liu1986relativistic,pennisi87}, and formalized a few years later by Geroch and Lindblom \cite{geroch1990dissipative}. An inequality like (\ref{entropycons}) only holds \textit{on-shell}, that is, when the conservative equations are satisfied. This requirement, together with the symmetry of $T^{ab}$, implies the local existence of a scalar $\chi$ which is a function of a covector $\xi_a$ and a symmetric traceless tensor $\xi_{ab}$, such that the conservative variables can be locally recovered as
\begin{eqnarray}
T^{ab} &=& \frac{\partial^{2}\chi}{\partial \xi_a \partial \xi_b} \label{Tfromchi} \, , \\
A^{abc}&=& \frac{\partial^{2} \chi}{\partial \xi_{a} \partial \xi_{bc}}  \label{Afromchi} - \frac{1}{4}\frac{\partial^2\chi}{\partial\xi_{a}\partial\xi_{de}}g_{de}g^{bc}.
\end{eqnarray}
Within this framework, the entropy current reads
\begin{equation}\label{eq:S_def}
S^a = \frac{\partial \chi}{\partial \xi_a} - T^{ab}\xi_b - A^{abc}\xi_{bc}\,,
\end{equation}
sourced by
\begin{equation}\label{eq:divS_def}
\sigma = -\xi_{ab}I^{ab}.
\end{equation}

This new set of variables $(\xi_a,\xi_{ab})$ comes out as ``Lagrange multipliers'' for the equations of motion, and we will refer to them as the \textit{abstract} variables. The key point is that all the information of the theory is now encoded into a single (sufficiently smooth) scalar field, $\chi(\xi_a,\xi_{ab})$, which has been found to be crucial for probing the symmetric-hyperbolicity of the evolution equations near equilibrium solutions \cite{Rubio18}. In fact, by introducing a collective abstract variable $\xi^A = (\xi_a,\, \xi_{ab})$, equations (\ref{Tabcons}) and (\ref{dissipeq}) can be set into the form
\begin{equation}\label{geroch-form-system}
\mathcal{K}^a{}_{AB}\nabla_a \xi^B = J_A,
\end{equation}
where
\begin{equation}
\mathcal{K}^a{}_{AB} := \frac{\partial^3 \chi}{\partial \xi_a \partial \xi^A \partial \xi^B}
\end{equation}
is the \textit{principal part} of (\ref{geroch-form-system}) (which by construction is symmetric in the capital indices) and $J_A := (0,0,\;I_{ab})$. The system is \textit{symmetric-hyperbolic} at $\xi^C$ provided there exists
a covector $\kappa_a$ such that the form $h_{AB}(\xi^C):=\kappa_a\mathcal{K}^a{}_{AB}|_{\xi^C}$ is \textit{positive-definite}; i.e., if $h_{AB}(\xi^C)\xi^A\xi^B > 0$ for any non-zero $\xi^A$. This algebraic condition guarantees that the theory admits a locally \textit{well-posed} initial-value formulation \cite{geroch1996partial}.

\subsection{Conformal invariance requirements}

From a phenomenological viewpoint, we aim to simulate ultra-relativistic fluids at high temperatures, taking into account dissipative effects and energy transport. In this regime, both kinetic and thermal energies are several orders of magnitude higher than the corresponding ``rest'' energy, thus becoming irrelevant\footnote{We refer the reader to the book \cite{Kremer02} for a complete microscopic description of this aspect, following a purely kinetic approach.}. Moreover, from dimensional arguments one can infer that there cannot be any intrinsic length scale for the theory, thus becoming it \textit{scale-invariant}. This symmetry means that the evolution equations should not change under any re-scaling of the background metric, for a proper re-scaling of the dynamical fields. More specifically, we say that the theory is  \textit{conformally invariant} if there exist \textit{conformal weights} $\alpha$ and $\beta$ such that, under a conformal transformation of the background metric, namely
\begin{equation}\label{Weyl-g}
g_{ab} \mapsto \Omega^2 g_{ab} \,,
\end{equation}
system (\ref{Tabcons} - \ref{dissipeq}) covariantly transforms under the map
\begin{equation}\label{AT-transf}
\left( 
\begin{array}{cc}
T^{ab} \\
A^{abc}
\end{array} 
\right)
\mapsto
\left( 
\begin{array}{cc}
\Omega^{\alpha}T^{ab} \\
\Omega^{\beta}A^{abc}
\end{array} 
\right).
\end{equation}
In fact, a transformation like (\ref{AT-transf}) leaves the equations unaltered if $\alpha = -6$ and $\beta=-8$ (in four spacetime dimensions).

As a consequence of this symmetry, two nontrivial conditions for the dynamical variables come out. Firstly, and as probably expected by the reader, the energy-momentum tensor $T^{ab}$ must be trace-free,
\begin{equation}\label{Tab-condition}
T^{ab}g_{ab} = 0.
\end{equation}
The above condition automatically fixes the equation of state (getting $\rho-3p=0$, where $\rho$ is the energy density and $p$ the fluid pressure). Furthermore, the constitutive tensor $A^{abc}$ must satisfy two extra algebraic relations, namely
\begin{eqnarray}\label{Acondition}
 \hat{n}_a A^{abc} -  \hat{n}_{a}A^{(bc)a} &=&0 \nonumber \\ 
 A^{abc}g_{ab}&=&0,
\end{eqnarray}
where $\hat{n}_{c}\equiv\nabla_{c} \Omega/\Omega$.
Both of them implies that $A^{abc}$ is symmetric in all its indices.

Under the above requirements, any conformal theory made up from a generating function which is quadratic in $\xi_{ab}$ (i.e., \textit{second order} theories for dissipative fluids) is uniquely parametrized by three free constants. Different choices of the parameters give rise to different conformally invariant theories, as will be shown in the next subsection. Furthermore, symmetric-hyperbolicity of the theory, which implies its well-posedness \cite{geroch1996partial}, further restricts the parameters. For second order theories, symmetric-hyperbolicity follows simply by requiring the parameter associated with the second-order contribution to be ``large enough'', as shown in \cite{Rubio18}.

\subsection{Generating function and conservative variables}

Conditions (\ref{entropycons}) and (\ref{eq:divS_def}) imply that the entropy production of the system depends purely on $\xi_{ab}$ and the source $I^{ab}$. Then, it becomes natural to associate $\xi_{ab}$ to the \textit{dissipative} degrees of freedom of the theory, and also to understand the generating function as an expansion in powers of $\xi_{ab}$ (aiming the different orders of dissipation). 

As mentioned above for conformally invariant theories, the most general scalar $\chi(\xi_a,\xi_{ab})$ up to second order in $\xi_{ab}$ reads
\begin{equation}\label{chiexpansion}
\chi = \chi^{0}(\mu) + \chi^{1}(\mu)\nu + \sum_{i=1}^{3}{\chi^{2}_i (\mu)\psi_i}\;, 
\end{equation}
where 
\begin{eqnarray}
\mu&\equiv&\xi^a\xi_a \nonumber \\ \nu&\equiv&\xi^{ab}\xi_a\xi_b\nonumber \\ \psi_{1}&\equiv&\xi^{ab}\xi_{ab}  \\ 
\psi_{2}&\equiv&\ell^a\ell_a\nonumber \\
\psi_{3}&\equiv&\nu^{2} \nonumber
\end{eqnarray}
and $\ell^a\equiv \xi^{ab}\xi_b$.
The functions $\chi^0(\mu)$, $\chi^1(\mu)$ and $\chi^2_i(\mu)$ are fixed by the conformal invariance conditions (\ref{Tab-condition}) and (\ref{Acondition}), getting 
\begin{eqnarray*}\label{sol-chi0-chi1}
\chi^0(\mu) &=& \frac{\chi_0}{\mu}\\
\chi^1(\mu) &=& \frac{\chi_1}{\mu^3}\\
\chi^2_i(\mu) &=& \frac{\chi_2}{\mu^{2+i}}\Theta_i
\end{eqnarray*}
where $\{\chi_j\}_{j=0}^{2}$ are three free parameters and
\[
\Theta_i \equiv \left\{
{\begin{array}{rcl}
1,&{\mbox{if}}& i = 1\\
-12,&{\mbox{if}}& i=2\\
24,&{\mbox{if}}& i = 3
\end{array}
}\right.
\]

The physical interpretation of the generating function (\ref{chiexpansion}) is the following. At zeroth order, $\chi^0(\mu)$ leads to the ideal conformal fluid, which satisfies the radiation equation of state ($\rho=3p$, with $\rho$ the energy density and $p$ the pressure). In local equilibrium, $\xi_{ab}=0$, and $\xi_a$ is a conformal Killing vector field (CKVF); i.e., $\nabla_{(a}\xi_{b)}=\zeta g_{ab}$, for some scalar $\zeta$. The set of CKVFs are called the \textit{equilibrium states} of the theory, and from them one can define the temperature of the fluid as $T\equiv 1/\sqrt{|\mu|}$. The stability of the equilibrium states is discussed in detail in \cite{geroch1990dissipative}. At first order, one gets the corresponding ``relativistic'' Navier-Stokes equations in the radiation regime \cite{Eckart40}. As it was shown in \cite{hiscock1983stability}, the theory up to this order presents generic instabilities when the high-frequency limit is reached. Interestingly, it was recently proposed an alternative first-order formulation for conformal fluids \cite{Bemfica21}, where the corresponding constitutive relations do not follow from an equation of the form (\ref{dissipeq}), but from a suitable gradient expansion. A numerical exploration of solutions of this theory was performed in 1+1 dimensions \cite{Pretorius21}, in 2+1 \cite{Figueras22} and in 3+1 \cite{Pretorius22}, and we aim to compare them with the simulations we will carry on throughout this work.

We recall the orthogonal decomposition of $\xi_{ab}$ in terms of the local equilibrium field $\xi_a$,
\begin{equation}\label{decomposition-xiab}
\xi_{ab} = \frac{4}{3}\frac{\nu}{\mu^2}\left(\xi_a\xi_b - \frac{\mu}{d}g_{ab}\right) + \frac{2}{\mu}\xi_{(a}r_{b)} + \tau_{ab},
\end{equation}
where 
\begin{eqnarray} \label{eq:r-tau-def}
    r_a &\equiv& \xi_{ab}\xi^b - \frac{\nu}{\mu}\xi_a\\
    \tau_{ab} &\equiv& h_{a}{}^ch_{b}{}^d\xi_{cd} + \frac{\nu}{3\mu}\left(g_{ab}-\frac{\xi_a\xi_b}{\mu}\right),
\end{eqnarray}
and $h_{a}{}^b$ is the projector onto the space orthogonal to $\xi_a$, namely $h_{a}{}^b=\delta_a{}^b-\xi_a\xi^b/\mu$.

Using the definitions (\ref{Tfromchi})-(\ref{Afromchi}), the full energy-momentum tensor $T^{ab}$ and constitutive tensor $A^{abc}$ can be written as the sum of the zeroth, first and second order contributions. For $T^{ab}$, we get
\begin{equation} \label{Tab-sum3}
    T^{ab} = T_0^{ab} + T_1^{ab} + T_2^{ab}
\end{equation}
where
\begin{eqnarray*}
    T_0^{ab} &=& \frac{-2\chi_0}{\mu^3}\left(\mu g^{ab}-4\xi^a \xi^b\right),\\
    T_1^{ab} &=&\chi_{1}\left[\frac{48\nu}{\mu^{5}}\xi^{a}\xi^{b}-\frac{24}{\mu^4}\xi^{(a}\ell^{b)}-\dfrac{6\nu}{\mu^{4}}g^{ab}+\dfrac{2}{\mu^{3}}\xi^{ab}\right]
\end{eqnarray*}
and
\begin{eqnarray*}
    T_2^{ab} &=& \frac{\chi_2}{\mu^5}\left[48\xi^{cd}\xi_{cd} - \frac{960\ell^c\ell_c}{\mu} + \frac{2880\nu^2}{\mu^2}\right]\xi^a\xi^b\\
    &+& \frac{\chi_2}{\mu^4}\left[-6\xi^{cd}\xi_{cd} + \frac{96\ell^c\ell_c}{\mu} - \frac{240\nu^2}{\mu^2}\right] g^{ab} \\
    &+& \frac{384\chi_2\xi^{(a}\xi^{b)c}\ell_c}{\mu^5} - \frac{1920\chi_2\nu\xi^{(a}\ell^{b)}}{\mu^6}
    + \frac{192\chi_2\ell^a\ell^b}{\mu^5} \\
    &+& \frac{96\chi_2\nu\xi^{ab}}{\mu^5} - \frac{24\chi_2\xi^{ac}\xi^b{}_c}{\mu^4}
\end{eqnarray*}
Similarly, for $A^{abc}$ we get
\begin{equation} \label{Aabc-sum2}
    A^{abc} = A_1^{abc} + A_2^{abc}
\end{equation}
where
\begin{eqnarray*}
A_1^{abc}&=&\frac{\chi_1}{\mu^3}\left[2g^{a(b}\xi^{c)}+g^{bc}\xi^{a}-\frac{6}{\mu}\xi^{a}\xi^{b}\xi^{c}\right]\\
A_2^{abc} &=&\dfrac{\chi_{2}}{\mu^3}
    \left[ \xi^{a}
        \left(\dfrac{-12}{\mu}\xi^{bc}+\dfrac{192}{\mu^2}
            \left(\xi^{(b}\ell^{c)}-\dfrac{\nu}{4}g^{bc}\right)\right.\right.\\
        &-&\left.\dfrac{480}{\mu^{3}}
            \left(\xi^{b}\xi^{c}-\dfrac{\mu}{4}g^{bc}\right)
        \nu\right)
    \\ 
    &-&\dfrac{24}{\mu}
        \left(g^{a(b}\ell^{c)}+\xi^{(b}\xi^{c)a}-\dfrac{1}{2}g^{bc}\ell^{a}\right)
    \\
    &+&\left. \dfrac{96}{\mu^{2}}
        \left(\ell^{a}
            \left(\xi^{b}\xi^{c}-\dfrac{\mu}{4}g^{bc}\right)
        +\nu
            \left(g^{a(b}\xi^{c)}-\dfrac{1}{4}\xi^{a}g^{bc}\right)
        \right)
    \right].
\end{eqnarray*}

From now on, we assume that $\xi^a$ is timelike (and so $\mu < 0$). Then, the normalized 4-vector $u^a=\xi^a/\sqrt{-\mu}$ is interpreted as the 4-velocity of the fluid, which allows to recast the typical form for the energy-momentum tensor, namely
\begin{equation}\label{fullTab}
    T^{ab} = \frac{4\rho}{3}\left(u^a u^b + \frac{g^{ab}}{4}\right) + 2u^{(a}Q^{b)} + \Sigma^{ab},
\end{equation}
being $\rho$ the energy density, $Q^a$ the heat flux and $\Sigma^{ab}$ the transverse traceless stress, up to second order in $\xi_{ab}$. In effect, they can be computed from (\ref{Tab-sum3}) by
\begin{eqnarray*}
\rho &=& T^{ab}u_au_b\\
Q^{a} &=& - h^{a}{}_bT^{bc}u_c\\
\Sigma^{ab} &=& h^{a}{}_ch^{b}{}_d T^{cd}.
\end{eqnarray*}
Then, by taking $\xi^a$ to be timelike and in order to ensure the energy density to be positive definite at every order, we require that $\chi_0<0$. Since the sign of $\chi_1$ is irrelevant as it is a global factor in $A^{abc}$, we take $\chi_1>0$. Finally, in order the principal part to be positive definite (an to ensure the symmetric-hyperbolicity), we require $\chi_2<0$.

\subsection{Fixing \texorpdfstring{$I^{ab}$}{Lg} with conformal weights}

We now give the most general expression of the source field $I^{ab}$, linear in $\xi_{ab}$, by requiring conformal invariance. Since the different contributions to the source field could covariantly transform in different ways under a metric re-scaling, we need to keep track of the conformal weights the different terms have. 

We say that a quantity $X$ has conformal weight $n$ if under the a local re-scaling for the metric
\begin{equation}
    \hat{g}_{ab} = \Omega^{2} g_{ab},
\end{equation}
such a quantity transforms as
\begin{equation}
    \hat{X} = \Omega^{-n} X
\end{equation}
We then write $\mathcal{CW}(X)=n$. It directly follows that $\mathcal{CW}(g_{ab})=-2$ and $\mathcal{CW}(g^{ab})=2$. Conformal invariance requirements implies straightforwardly that $\mathcal{CW}(\nabla_a T^{ab})=6$ and $\mathcal{CW}(\nabla_a A^{abc})=8$. Also, we get that $\mathcal{CW}(I^{ab}) = 8$ and $\mathcal{CW}(I_{ab})=4$.
Finally, one also has that $\mathcal{CW}(\xi_a) = -2$ and $\mathcal{CW}(\xi_{ab})=-4$, which implies that $\mathcal{CW}(\xi^a) = \mathcal{CW}(\xi^{ab})=0$. With this information, we can get the corresponding conformal weights for each of the terms that constitute the source tensor. In fact, a decomposition of $I^{ab}$ analog to the one given in (\ref{decomposition-xiab}) for $\xi_{ab}$ also follows, leading to
\begin{equation}
    I^{ab} = I_0 \nu \left(\xi^a \xi^b-\frac{\mu}{d}g^{ab}\right) + I_1 \xi^{(a}r^{b)} + I_2\tau^{ab}
\end{equation}
where the $I_j$ are powers of $\mu$ to be fixed, up to a constant factor, from conformal invariance requirements. Firstly, we get that
\begin{equation}
    \mathcal{CW}(I^{ab})=\mathcal{CW}(\nabla_a A^{abc}) = 8.
\end{equation}
Then, since $\mathcal{CW}(\xi^a)=0$ and $\mathcal{CW}(\nu)=-4$, it must be $\mathcal{CW}(I_0(\mu)) = 12$. Using that $\mathcal{CW}(\mu)=-2$ and for $p\in\mathbb{R}$, $\mathcal{CW}(\mu^{p})=-2p$, we get that
\begin{equation}\label{I0}
    I_0 = -\frac{C_0}{\mu^{6}},
\end{equation}
for some real constant $C_0$.
Analogously, we get
\begin{eqnarray}\label{I1}
    I_1 &=& \frac{C_1}{\mu^{5}},
\end{eqnarray}
\begin{eqnarray}\label{I2}
I_2 &=& -\frac{C_2}{\mu^{4}},
\end{eqnarray}
for some real constants $C_1$ and $C_2$, where we have also used that $\mathcal{CW}(r^a) = -2$ and $\mathcal{CW}(\tau^{ab})=0$.
Finally, the source field is fixed as
\begin{equation}\label{fuente}
    I^{ab} = - \frac{C_0\nu}{\mu^{6}}\left(\xi^a \xi^b-\frac{\mu}{4}g^{ab}\right) + \frac{C_1}{\mu^{5}} \xi^{(a}r^{b)} - \frac{C_2}{\mu^{4}}\tau^{ab}.
\end{equation}
With the above expression for the source, the entropy production (\ref{eq:divS_def}) reads
\begin{equation}\label{sigma-explicit}
    \sigma=C_{0}\dfrac{\nu^{2}}{\mu^{6}} - \dfrac{C_{1}}{\mu^{5}}r^{a}r_{a} + \dfrac{C_{2}}{\mu^{4}}\tau^{ab}\tau_{ab}.
\end{equation}
Therefore, since $\mu<0$, we get that $\sigma>0$ if and only if $C_i \geq 0$, with at least one of them strictly positive.

\subsection{Equation structure at equilibrium and free parameters}

Since the dynamical fields are obtained as  derivatives of a generating function, they are defined up to an overall constant, for which we can set $\chi_0 = -1$. This choice only changes the energy-momentum conservation equation, as the corresponding term of the generating function depends only on $\xi_a$. Also, we choose the negative sign so that $T^{00}$ is non-negative.
The term proportional to $\chi_1$ is linear on  the dissipative variables, $\xi_{ab}$, so we can re-scale those variables to set $\chi_1=1$. The remaining constants are $\chi_2$ and $C_i$, the three parameters in the source field (\ref{fuente}). With all this, we found it useful to make the following rescaling
\begin{equation}
\chi_2 \to -\frac{\chi_2 \chi_0}{\chi_1^2}
\end{equation}
and 
\begin{equation}
C_i \to -\frac{C_i\;\chi_0}{\chi_1^2}.
\end{equation}
Finally, we recall that in order for the theory to be hyperbolic near equilibrium states, the absolute value of $\chi_2$ must be chosen to be large enough \cite{Rubio18}. 

A physical interpretation of the free constants in the theory can be obtained by analyzing the general structure of the dynamical equations at equilibrium.
In this regime, we have $\xi_{ab}=0$, and therefore there is no entropy production (see eq. (\ref{entropycons})). The equations (\ref{Tabcons})-(\ref{dissipeq}) remain
\begin{align}
\nabla_a T_0^{ab} &= 0 \\
\nabla_a A_1^{abc} & = 0
\end{align}

If $\chi_1 \neq 0$, they imply that $\xi_a$ is a conformal Killing vector field (provided the metric admits one). 
If $\chi_1 = 0$ only the first equation survives, namely the relativistic Euler's equations for radiation. Perturbations travel to the speed of sound for this case, namely $v^s_{\pm} = \pm\sqrt{1/3}$.

Now, if we allow generic perturbations \textit{off-equilibrium} (i.e., $\xi_a = \xi^0_a + \delta \xi_a$, $\xi_{ab} = \delta \xi_{ab}$). Then in the case $\chi_1=0$ we obtain the decoupled system 
\begin{align}
\nabla_a T_0^{ab} &= 0 \\
\nabla_a A_2^{abc} & = I^{bc}.
\end{align}

The perturbation of the off-equilibrium quantities give rise to other three propagation speeds: the {second speed of sound} $v^{ss}_{\pm} = \pm \sqrt{3/5}$, and a standing (or \textit{zero}) mode, $v^{s}_0 = 0$.
The existence of a zero mode can be guessed since, when $I^{bc} = 0$, the system is time symmetric. Thus there are as many positive roots as negative. So an even number of modes means that at least one of them must have zero speed. The negative definite character of $I^{ab}$ implies that those modes will also have a decay rate. If we turn on the interaction, i.e., if we set $\chi_1 \neq 0$, then a general theorem for these types of theories states that all modes will acquire a non-zero decay rate (see \cite{Kreiss1997GlobalEA} for details).

In the limit $\chi_2 \to 0$ and at equilibrium\footnote{Notice that such a limit is only formal, since when $\chi_2\to 0$, the system ceases to be well-posed, and generic instabilities are present in an arbitrary frame (see \cite{hiscock1983stability} for details).}, we can write the system as,
\begin{align}
\nabla_a T_0^{ab} + \nabla_a T_1^{ab} &= \frac{\partial T_0^{ab}}{\partial \xi_c} \nabla_a \xi_c + \frac{\partial T_1^{ab}}{\partial \xi_{cd}} \nabla_a \xi_{cd} = 0 \label{eq:par}\\
\nabla_a A_2^{abc} & = \frac{\partial A_1^{abc}}{\partial \xi_d} \nabla_a \xi_d = I^{bc} = M^{bcpq} \xi_{pq}, \nonumber
\end{align}
where $M^{bcpq}$ is a negative definite matrix that only depends on the equilibrium variables. Thus, we can invert the relation and get the so called \textit{constitutive relations},
\begin{equation}
\xi_{pq} = M^{-1}_{pqbc} \frac{\partial A_1^{abc}}{\partial \xi_d} \nabla_a \xi_d
\end{equation}
Plugging them into eq. (\ref{eq:par}) we get a \textit{parabolic-like} equation with a diffusion time-scale
\begin{equation}
\tau_d\sim\frac{\chi_1^2}{C_i \chi_0}.
\end{equation}

Finally, in the limit of very large values for $\chi_2$, but still for very small initial values of $\xi_{ab}$, (that is, $\chi_2 \xi_{ab}$ finite) the \textit{constitutive relations} do not matter much and the system behaves as having very little dissipation with its equilibrium sector behaving as Euler's system. That is, the relevant equations in this regime would be:
\begin{align}
\nabla_a T_0^{ab}  &= 0 \\
\nabla_a A_2^{abc} &= 0.
\end{align}
Thus, we would find the same propagation speeds as before, but with no dissipation.

\section{Numerical Implementation}
\label{sec-3}

\subsection{Evolution variables and (3+1)-decomposition}

Although the abstract variables $(\xi_a,\xi_{ab})$ are the natural ones for probing the symmetric-hyperbolicity of the theory, they are not suitable for a numerical implementation. The reason is that the evolution equations in abstract variables are not in conservative form, thus being not possible to capture shock formation. For this, we get back to the original formulation in terms of the conservative variables $(T^{ab}, A^{abc})$, and follow a fully conservative scheme. Given the nonlinear relation between both set of variables, this procedure will require an iterative inversion of the map that gives the conservative variables in terms of the abstract ones, in order to reconstruct the corresponding fluxes.

To find an evolution system of equations, we proceed as follows. Having fixed the spacetime background metric as to be Minkowski, $\eta_{ab}$, we first pick a spatial hypersurface $\Sigma$ and put global inertial coordinates $(x,y,z)$ on it\footnote{The generalization to more general background metrics and coodinate systems is straightforward.}. Then, we consider a vector field $t^a$ which is transverse to $\Sigma$ (we actually choose it to be everywhere orthogonal), and extend the coordinates $(x,y,z)$ in a way that they are constant along the integral curves of $t^a$. We take as ``time coordinate'' the function $t:\mathcal{M}\to\mathbb{R}$ which is zero on $\Sigma$ and satisfies $t^a\nabla_a t = 1$, where $\nabla$ is the connection compatible with $\eta_{ab}$. Then, $t^a = (\partial/\partial t)^a$, and the fluid 4-velocity reads
\begin{equation}
    u^a = \gamma (1,\, v^i),
\end{equation}
where $v^i$ is the spatial 3-velocity of the fluid and $\gamma = 1/\sqrt{1-v^iv_i}$ is the Lorentz factor.

The conservation equations (\ref{Tabcons})-(\ref{dissipeq}) become, then,
\begin{eqnarray}
\partial_t T^{00} &=& -\partial_i T^{0i}, \\
\partial_t T^{0i} &=& -\partial_j T^{ij}, \\
\partial_t A^{000} &=& -\partial_i A^{00i} + I^{00} \label{eq-A1}\\
\partial_t A^{00i} &=& -\partial_j A^{0ij} + I^{0i} \label{eq-A2}\\
\partial_t A^{0ij} &=& -\partial_k A^{ijk} + I^{ij}\,, \label{eq-A3}
\end{eqnarray}
together with the trace-free condition
\begin{equation}
    \eta_{ab}A^{abc}=0,
\end{equation}
which must be checked at each time step.

The problem reduces then to obtain expressions for the fluxes $T^{ij}$ and $A^{ijk}$ in terms of the evolution variables\footnote{The rest of the components can be computed using the full symmetry of $T^{ab}$ and $A^{abc}$.} $\{T^{00},T^{0i},A^{000}, A^{00i},A^{0ij}\}$. Since these expressions are defined in terms of the abstract variables (see expressions (\ref{Tab-sum3})-(\ref{Aabc-sum2})), we need to express the abstract variables in term of the conserved quantities we are evolving, namely invert the relation among $\{T^{00},T^{0i},A^{000}, A^{00i},A^{0ij}\}$ and $\{\xi_a, \xi_{ab}\}$, from definitions (\ref{Tab-sum3})-(\ref{Aabc-sum2}). This is in general not a simple task, since there is not a closed form for expressing it~\footnote{This is not surprising, as this is also the case in relativistic MHD (see for instance \cite{Font00})}. Thus, at each time step, and for each grid point, we need to numerically invert this relation. We use a Newton-Raphson method for this. Explicitly, the inversion we are seeking is among: \begin{equation}{\label{mapa-inv}}
\left(
\begin{array}{c}
T^{00} \\
T^{0i} \\
A^{000} \\
A^{00i} \\
A^{0ij} 
\end{array}
\right)
\to
\left(
\begin{array}{c}
\mu \\
v^i \\
\nu \\
r^i \\
\tau^{ij} 
\end{array}
\right),
\end{equation}
where $\nu=\xi^{ab}\xi_a\xi_b$, and $r_a$ and $\tau_{ab}$ have been introduced in Eqs. (\ref{eq:r-tau-def}).

We now apply this formulation to an effective one-dimensional case and find numerical solutions to the corresponding system of equations.

\subsection{Evolution equations with planar symmetry}

In order to simplify the numerical implementation as a first exploration of the dynamics of this family of theories, we look for configurations which are plane-symmetric (also known as ``slab'' symmetric) in Minkowski spacetime, $g_{ab}=\text{diag}(-1,1,1,1)$, and consider the effective 1+1 evolution system. For doing so, we take cartesian coordinates $(t,x,y,z)$ and consider the flow dynamics over the $x$ axis. The 4-velocity  of the fluid reads $u^a=\gamma(1,v,0,0)$, where $\gamma=(1-v^2)^{-1/2}$ and the evolution equations reduce to
\begin{eqnarray} \label{1+1-evo-eqs}
\partial_t T^{00} &=& -\partial_x T^{01}, \nonumber\\
\partial_t T^{01} &=& -\partial_x T^{11}(T^{00},T^{01},A^{000},A^{001},A^{011}), \nonumber\\
\partial_t A^{000} &=& -\partial_x A^{001} + I_o \label{system-slab}\\
\partial_t A^{001} &=& -\partial_x A^{011} + I_1 \nonumber\\
\partial_t A^{011} &=& -\partial_x A^{111}(T^{00},T^{01},A^{000},A^{001},A^{011}) + I_2 \nonumber
\end{eqnarray}

The evolution variables are the energy and momentum densities, and three of the components of the constitutive tensor $A^{abc}$, namely $(T^{00}, T^{01}, A^{000},A^{001},A^{011})$. Instead, the fluxes
$(T^{11},A^{111})$ are expressed in terms of $(\mu,v^1,\nu, r^1, \tau^{11})$ which are the relations we invert:

\begin{equation}{\label{mapa-inv-2}}
\left(
\begin{array}{c}
T^{00} \\
T^{01} \\
A^{000} \\
A^{001} \\
A^{011} 
\end{array}
\right)
\to
\left(
\begin{array}{c}
\mu \\
v \\
\nu \\
r^1 \\
\tau^{11} 
\end{array}
\right).
\end{equation}
The explicit map between these variables can be found in Appendix \ref{app-expeqs}.

The extra abstract variables are obtained, using that $r^au_a=0$, to get that $r^0=v r^1$, and 
\begin{equation}
    \tau^{ab}u_b=0, \quad \tau^{ab}g_{ab}=0, \quad \tau^{22}=\tau^{33},
\end{equation}
to get,
\begin{eqnarray}
    \tau^{00} &=& v^2\tau^{11} \nonumber\\ \tau^{01}&=&v\tau^{11}\\ \tau^{22}&=&\tau^{33}=\frac{1}{2}(v^2-1)\tau^{11}. \nonumber
\end{eqnarray}

The Jacobian of the transformation between conservative and abstract variables was obtained symbolically from the expressions for the fluxes used in the code and then automatically converted into efficient matrix functions for using in the numerical part of the code.

The inversion from conservative to primitive variables was only used to find the non trivial fluxes and the source terms. For the other equations, the conservative variables were directly used, making the code simpler and with fewer computations. 
The sources are $I_0:=I^{00}$, $I_1:=I^{01}$ and $I_2:=I^{11}$, whose components are taken directly from Eq. (\ref{fuente}).

Thus, the relevant components of the abstract variables which are computed are $\{v,\mu,\nu,r^1,\tau^{11}\}$, for which we will refer them as the \textit{fluid} variables. In order to get them in terms of the conservative ones, we numerically invert expressions (\ref{Tab-sum3})-(\ref{Aabc-sum2}) and, after that, we compute the nontrivial fluxes.

By sweeping on the values for $\chi_2$, we found that the Jacobian of this transformation turns out to be singular for a particular value of $\chi_2$, which is $\chi_2=-5/48$. This curious fact can be understood from a simple procedure further detailed in Appendix (\ref{cuenta-pablo}). In the following subsection, we show how to use the fluid variables in order to assess the equations implemented in the code. 

\subsection{Characteristic structure}
\label{sec:charspeed}

In order to check that the evolution equations (\ref{1+1-evo-eqs}) are self-consistent and well-implemented throughout the code, we transform them to symbolic equations. This means that we work with the equations as mathematical expressions and operate over them using computer algebra, which allows us to calculate derivatives of the expressions with respect to its arguments. This calculations where done using the Julia package \textit{Symbolics.jl} \cite{gowda2021high}. Using this we can assess the symmetry of the principal part. 
System (\ref{1+1-evo-eqs}) has the form 
\begin{equation}
    \partial_t \mathbf{c}^j = \partial_{x} F^{j}(\mathbf{c}, \mathbf{f}) + I^j(\mathbf{c}, \mathbf{f}),
\end{equation}
where $\mathbf{c}^j=\{T^{00}, T^{01}, A^{000}, A^{001}, A^{011}\}$ are the conservative variables and $\mathbf{f}^j=\{\mu,v,\nu,r_1,\tau_{11}\}$ the fluid ones. The Jacobian of the flux function with respect to the conservative variables is given by
\begin{equation}
\frac{d F^{\ell}}{d \mathbf{c}^j} = \frac{\partial F^{\ell}}{\partial \mathbf{c}^j} +\frac{\partial F^{\ell}}{\partial \mathbf{f}^k}\frac{\partial \mathbf{f}^k}{\partial \mathbf{c}^j},
\label{eq:jacobian-PC}
\end{equation}
whose eigenvalues give us the propagation speeds of the system.

From the equations (\ref{Tfromchi}) and (\ref{Afromchi}), it is easy to see that the matrix 
\begin{equation}
M^{\ell}{}_{A}:=\frac{d F^{\ell}}{d \mathbf{c}^j}\dfrac{d\mathbf{c}^j}{d\xi_A}
\end{equation}
is symmetric, with $\xi_A=\{\xi_0,\xi_1,\xi_{00},\xi_{01},\xi_{11}\}$. Since $\xi_{ab}$ is traceless, we need to substract the trace whenever deriving with respect to it. Besides, since $\xi_{01} = \xi_{10}$, in one must divide the derivatives with respect to $\xi_{01}$ by $2$ in the scheme. In order to guarantee these subtleties, we multiply $M^{\ell}{}_{A}$ by the matrix
\begin{equation}
S = \begin{pmatrix} 
1 &0 &0& 0& 0 \\
     0& 1& 0& 0& 0\\
     0& 0& \frac{3}{4}& 0 &\frac{1}{4}\\
     0& 0& 0& \frac{1}{2}& 0\\
     0& 0& \frac{1}{4}& 0 &\frac{3}{4}\\
\end{pmatrix}
.
\end{equation}
We computed all these matrices using the Julia package \textit{Symbolics.jl} \cite{gowda2021high}, and checked that this symmetry is present in the equations in our code.

\begin{figure}
    \centering
    \includegraphics[width=\linewidth]{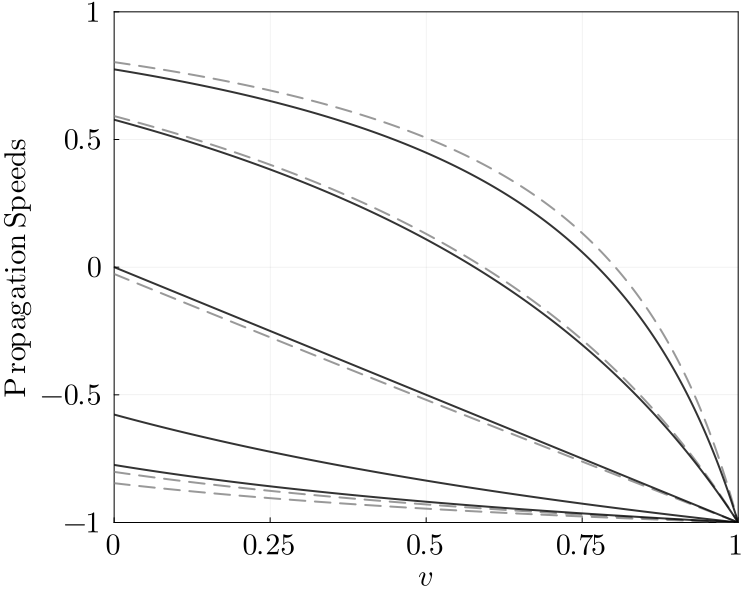}
    \caption{Propagation speeds with respect to the fluid velocity. The continuous line corresponds to $\nu = r_1 = \tau_{11} = 0$, while the dashed line corresponds to $\nu = r_1 = \tau_{11} = 0.1$ at $v = 0.0$. Notice that the values of $ r_1$ and $\tau_{11}$ change according to a Lorentz transformation when $v \neq 0$.}
    \label{fig:propagation_speeds}
\end{figure}

Equation (\ref{eq:jacobian-PC}) also allows for the calculation for each particular choice of the variables $\mathbf{c}$ and $\mathbf{f}$. In particular, by setting $\mathbf{f} = (\mu, 0, 0, 0, 0)$, we recover the well-known propagation speeds of an ideal, pure-radiation fluid, namely $v_o=0$ (the standing mode), $v_1^{\pm}=v\pm\sqrt{1/3}$ (first sound speed) and furthermore $v_2^{\pm}=\pm\sqrt{3/5}$ (a second sound speed characteristic of hyperbolic dissipative systems). If we now move to a frame where the fluid velocity is non-zero, we would expect the propagation speeds to decrease and go to $-1$ when $v\rightarrow 1$. This is actually the case shown in figure \ref{fig:propagation_speeds}.
Since $r_1$ is a component of a vector field, and the same for $\tau_{11}$ as a component of a tensor field, they should properly transform under a Lorentz boost. We assessed this behaviour in the code, and fully validate the consistency of the evolution equations implemented.

\subsection{Numerical method}
\label{num-method}

We consider the one-dimensional domain from $x=-L$ to $x=L$ and uniformly discretize it with $N$ grid points, such that the spatial step is $\Delta x = 2L/(N-1)$ and the grid points are $x_i=-L+i\Delta x$, $i=0,\cdots,N-1$. Then, for any time $t$, we approximate the value of a given function $u(t,x)$ over the numerical domain as $u_i(t):=u(t,x_i)$. We also set periodic boundary conditions.

As stated before, the evolution equations have the general form
\begin{equation}
    \partial_t u + \partial_x F(u) = g(u),
\end{equation}
where $F$ are the fluxes and $g$ is the source term. Then, our spatial discretization for the equations is given by a finite difference scheme, which takes the form
\begin{equation}
    \frac{\mathrm{d}u_i(t)}{\mathrm{d}t} + \frac{\hat{F}_{i+\frac{1}{2}} - \hat{F}_{i-\frac{1}{2}}}{\Delta x}  = g(u_i(t)),
\end{equation}
where $\hat{F}_{i+\frac{1}{2}} := \hat{F}(u_{i-p},...,u_{i+q})$ is a consistent numerical flux satisfying $\hat{F}(u,...,u) = F(u)$, and $p$ and $q$ depend on the chosen numerical method. In our code, these fluxes are reconstructed using the WENO-Z scheme. Given a function $h(u)$, the WENO schemes allow to reconstruct the approximate value of $h$ at the half points $x_{i+\frac{1}{2}}$ by using either a left biased weighted combination of the cell averages $\bar{h}_{j-2}$, $\bar{h}_{j-1}$, $\bar{h}_{j}$, $\bar{h}_{j+1}$ and $\bar{h}_{j+2}$ or a right biased  combination of the cell averages $\bar{h}_{j-}$, $\bar{h}_{j}$, $\bar{h}_{j+1}$, $\bar{h}_{j+2}$ and $\bar{h}_{j+3}$. The weights for each coefficient depends on the smoothness of the numerical solution, and are chosen so that the approximation is fifth order accurate when no discontinuity is present, and third order accurate in the neighbourhood of a shock. The difference between different WENO schemes lies on the way the weights are chosen. In particular, we chose the WENO-Z from the set of WENO schemes as it is the most accurate (having also tried with other central schemes, as the MP5 and Kurganov-Tadmor ones). Further details on the WENO-Z algorithm used in our code are given in Appendix \ref{app-wenoz}. We refer the reader to the article by Shu \cite{doi:10.1137/070679065} for more information about WENO schemes for the solution of conservation laws.

\begin{figure}
    \centering
    \includegraphics[width=\linewidth]{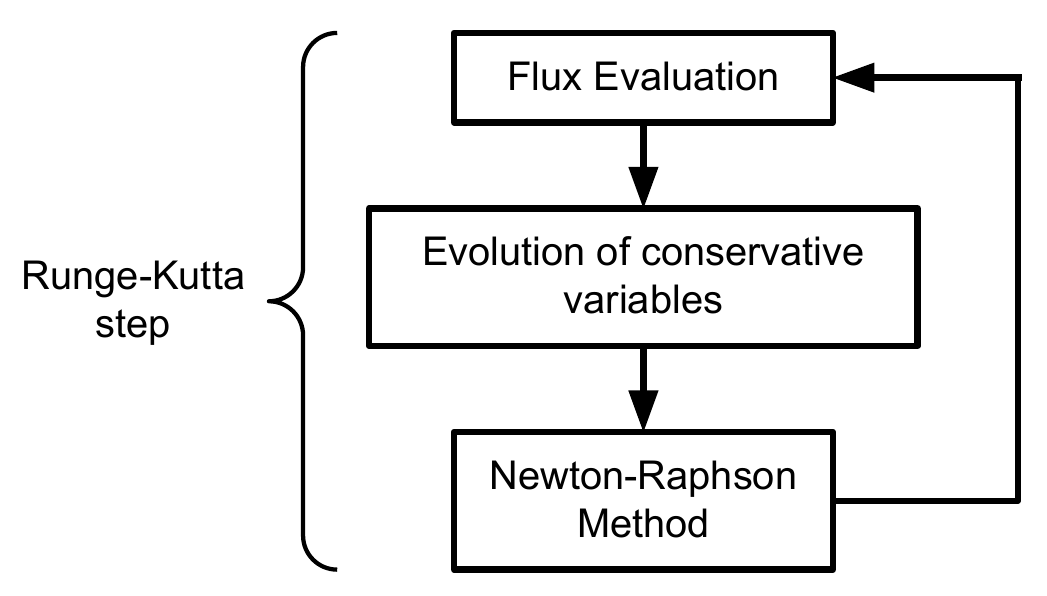}
    \caption{Description of the algorithm inside each Runge-Kutta internal step.}
    \label{fig:single-RK-step}
\end{figure}

For the purpose of stability we must ensure correct upwinding, We achieve this by using the Lax-Friedrichs flux splitting, in which the flux is decomposed in two parts,
\begin{eqnarray}
    F(u) = F^{+}(u)+F^{-}(u),
\end{eqnarray}
\begin{equation}
    F^{\pm}(u) = \dfrac{1}{2}(F(u)\pm\alpha u),
\end{equation}
where $\alpha:=\max_{u}|F'(u)|$ is the maximum propagation speed of the system. This way the propagation speeds are positive for $F^{+}$ and negative for $F^{-}$. Once the flux is split, we can calculate $\hat{F}^{+}_{i+\frac{1}{2}}$ using a left biased WENO-Z reconstruction and $\hat{F}^{-}_{i+\frac{1}{2}}$ using a right biased WENO-Z reconstruction, and calculate $F_{i+\frac{1}{2}} = F^{+}_{i+\frac{1}{2}} + F^{-}_{i+\frac{1}{2}}$.
For the time evolution, we define a time step $\Delta t$ so that $u_{i}^{n} := u_{i}(n\Delta t)$, and implement a third-order accurate strong stability preserving (SSP) TVD (Total Variation Diminishing) Runge-Kutta scheme (also known as SSPRK33), which is appropriate for essentially non-oscillatory shock capturing schemes (see \cite{SHU1988439} and references therein). All the algorithm was implemented in \url{Julia} \url{1.7.2}.
Once we have evolved for the conservative variables for a time step (internal to the RK) we invert the relations given in (\ref{Tab-sum3})-(\ref{Aabc-sum2}) and compute the corresponding fluid variables. From them, the fluxes $F_1$ and $F_2$ are obtained and so the whole right-hand-side of the evolution equations in order to complete the cycle. For the variables inversion, we implemented the Newton-Raphson method, using the results from the previous time step as a seed. The Jacobian of the transformation was obtained from the \textit{Symbolics.jl} package provided by Julia \cite{gowda2021high}, which allowed us to obtain both an analytic expression for it, and an efficient Julia numerical function. After doing this inversion, we evaluated the flux function and therefore evolved the evolution variables. This was then repeated for each step until the desired final time of integration was reached. A diagram of this scheme can be seen in figure \ref{fig:single-RK-step}.

\section{Results}
\label{sec-4}

In this section we present numerical results of the conformal theory introduced in \cite{Rubio18}. We focus the discussion in the evolution from two different initial data for the energy density: (i) a smooth gaussian profile and (ii) a discontinuous profile. We also set initially the dissipative variables to zero and see how they evolve due to the conservation equations, and compare the resulting dynamics with the one corresponding to the perfect fluid (Euler's equations).

\subsection{Initial data}

Keeping in mind the discussion about the free parameter space of the theory, we take $\chi_0=-1$ and $\chi_1=1$, and keep free $\chi_2<0$, as well as the three parameters of the source term, $C_i$, introduced in (\ref{fuente}). In order to explore how the dynamics is modified when changing the parameters, we consider two different initial data. Since we are interested in comparing the solutions with the case of the perfect fluid (which follow the Euler equations with a pure radiation equation of state), we set $\nu(0,x) = r_1(0,x) = \tau_{11}(0,x) = 0$. We also set $v(0,x)=0$ for the fluid initial velocity and consider a static initial data.

In order to compare the obtained results with a previous work \cite{Pretorius21}, for the smooth initial configuration we consider the following profile for the energy density:
\begin{equation} \label{eq:gaussian-initial-data}
    T^{00}(0,x) = Ae^{-x^2/\omega^2}+\delta, 
\end{equation}
with $x\in(-100,100)$, and setting $A = 0.4$, $\omega = 5$ and $\delta = 0.1$. Notice that since both the initial velocity profile and all the dissipative variables are initially set to zero, the data (\ref{eq:gaussian-initial-data}) corresponds also to the initial internal energy of the fluid. Also, for the discontinuous initial data we choose
\begin{equation}\label{eq:step-initial-data}
    T^{00}(0,x)=(\epsilon_{\mbox{\tiny{R}}}-\epsilon_{\mbox{\tiny{L}}})\dfrac{\mbox{erf}(x)+1}{2}+\epsilon_{\mbox{\tiny{L}}},
\end{equation}
where $\mbox{erf}(x)$ is the standard error function
\begin{equation}
    \mbox{erf}(x)=\frac{2}{\sqrt{\pi}}\int_{0}^{x}{e^{-t^2}\mbox{d}t}
\end{equation}
and the left and right parameters are set to $\epsilon_{\mbox{\tiny{L}}} = 0.4$ and $\epsilon_{\mbox{\tiny{R}}} = 0.1$.

Finally, we set the initial data by giving first the fluid variables. From the formula for the energy density of a perfect fluid $T^{00} = 6\chi_0\mu^{-2}$, we solve for $\mu$, and then we get the corresponding values of the conservative variables. Notice that even if the dissipative variables are equal to zero, this does not mean that $A^{000}$, $A^{001}$ and $A^{011}$ will also be zero.

\subsection{Parameter sweep}

\subsubsection{Effect of \texorpdfstring{$\chi_2$}{Lg}}

\begin{figure*}
    \centering
    \includegraphics[width =0.9\textwidth]{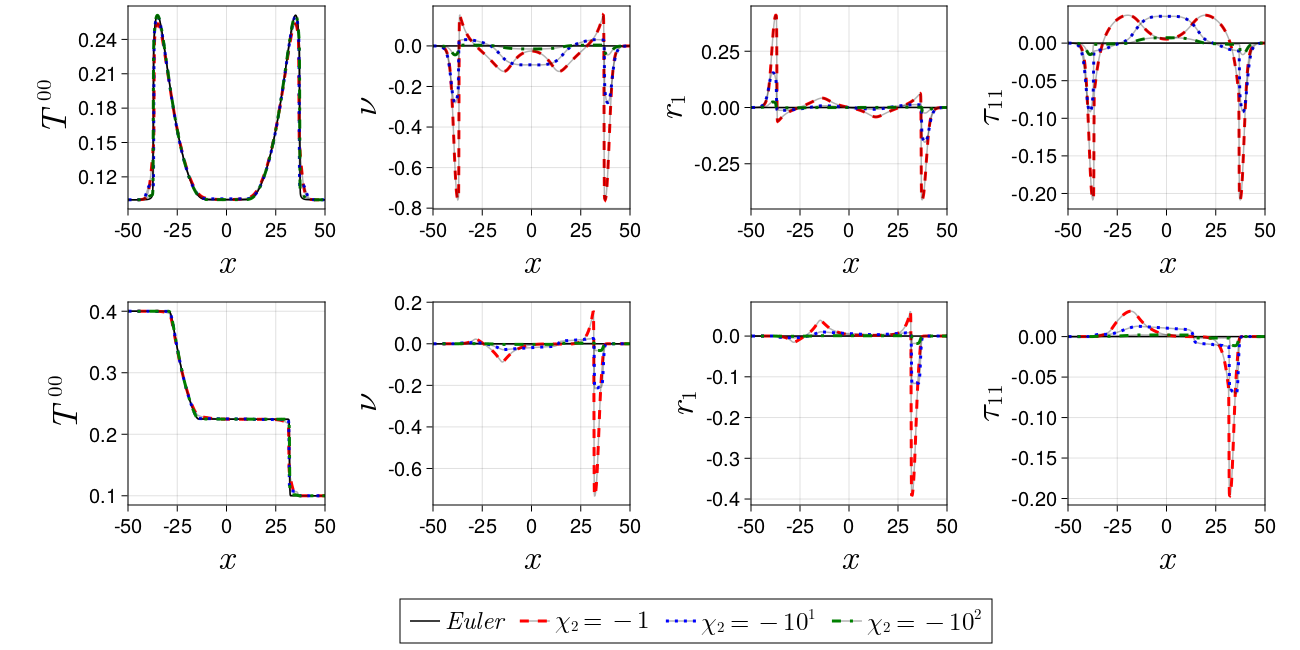}
    \caption{Energy density (gaussian profile at the top panel; discontinuous profile at the bottom panel) and the three dissipative variable profiles at $t=47\,\mbox{GeV}^{-1}$ for different values of $\chi_2$. We set $C_0 = C_1 = C_2 = 10$ for the parameters of the source.}
    \label{fig:chichonstep-chi2var-Enur1tau11}
\end{figure*}

We first analyze the effect of the variable $\chi_2$ on the energy and the dissipative variables. We do this by changing $\chi_2$ restricted in the range $(-1000,-1)$, and by setting $C_0 = C_1 = C_2 = 10$. As a reference, we also evolve the Euler equations, whose solutions can be achieved by setting $\chi_1 = 0$, which will always allow to evolve a perfect fluid as long as the dissipative variables are set initially to zero. Results can be seen in figure \ref{fig:chichonstep-chi2var-Enur1tau11}.
We notice a greater amplitude of the dissipative variables for smaller values of $|\chi_2|$, and a slight difference in the energy density. For the gaussian profile, a smaller second peak can be seen, which is moving faster than the rest, while the shock in the discontinuous initial data seems to be slightly smoother. These effects are expected given that the system has now a second sound speed, and therefore new modes propagating faster than the ones expected from the Euler's equations.

From what we learnt in this parameter exploration, and in order to observe a significant effect on the dissipative variables of the system, we choose to fix $\chi_2 = -1$ for all subsequent simulations. Higher values of $\chi_2$ are then ignored since the solution approximates the Euler system in the limit $\chi_2 \rightarrow -\infty$.

\subsubsection{Effect of \texorpdfstring{$C_0$}{Lg}, \texorpdfstring{$C_1$}{Lg} and \texorpdfstring{$C_2$}{Lg}}

We now focus our attention on the individual contributions of $C_0$, $C_1$ and $C_2$, the three free parameters of the source term (\ref{fuente}). To analyze each one separately, we  vary one of them independently, and fix the remaining two parameters to a high value, in order not to take into account their contribution in the subsequent dynamics.
\begin{figure*}
    \centering
    \includegraphics[width =0.9\linewidth]{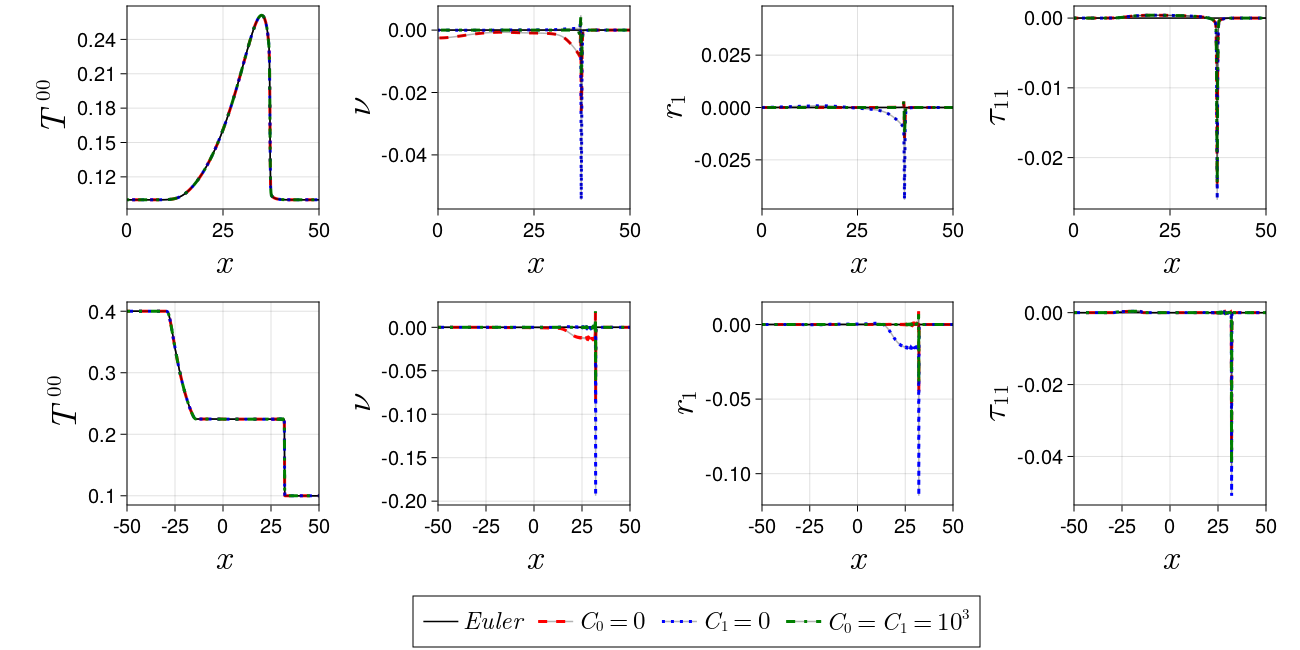}
    \caption{Energy density (gaussian profile at the top panel; discontinuous profile at the bottom panel) and the three dissipative variable profiles at $t=47\,\mbox{GeV}^{-1}$ for different values of $C_0$ and $C_1$. Unless specified, all other $C_i's$ are fixed to $10^3$.}
    \label{fig:chichon-C0C1var-Enur1tau11}
\end{figure*}
Heuristically, one would expect the system to work in this way, as for a given large value for $C_i$, the dissipative variables that are multiplied by $C_i$ must be ``small'', in order to compensate the right-hand side of the equation. Thus, since the dissipative variables are ``small'', their contribution to $T^{ab}$ is small too, and so the evolution becomes closer to the Euler system. By performing several runs with different large values for $C_i$, we saw that a ``large-enough'' value in order to \textit{turn off} the contribution of the corresponding terms in the source is $C^{\text{\tiny{off}}}_i=10^3$.
We proceed then to evolve the system by modifying only one of the $C_i$'s in a range from $0$ to $10^2$, setting down the other two variables in $C^{\text{\tiny{off}}}_{k\neq i} = 10^3$. We found that neither $C_0$ nor $C_1$ have a significant effect on the energy density function, as well as the dissipative variables remain almost unchanged. This behaviour can be seen in Figure \ref{fig:chichon-C0C1var-Enur1tau11}.
\begin{figure*}
    \centering
    \includegraphics[width =0.9\linewidth]{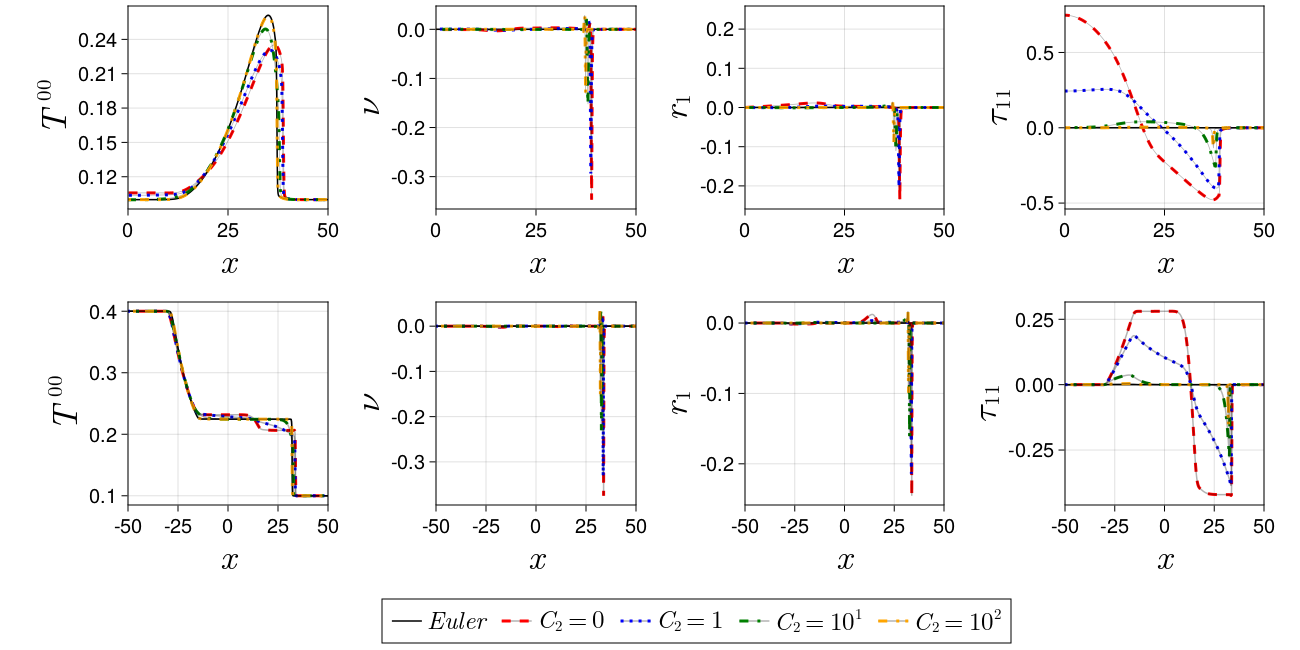}
    \caption{Energy density (gaussian profile at the top panel; discontinuous profile at the bottom panel) and the three dissipative variable profiles at $t=47\,\mbox{GeV}^{-1}$ for different values of $C_2$ and with $\chi_2 = -1$, $C_0 = C_1 = 10^3$.}
    \label{fig:chichonstep-C2var-Enur1tau11}
\end{figure*}
Even though a smaller $C_0$ increases the magnitude of $\nu$ and a smaller $C_1$ increases the magnitude of $r_1$, there is no discernible difference on neither $T^{00}$ nor $\tau_{11}$.

On the other hand, Figure  \ref{fig:chichonstep-C2var-Enur1tau11} shows that only modifying $C_2$ results in a noticeable change in the energy density and dissipative variables distribution.
\begin{figure*}
    \centering
    \includegraphics[width =0.9\linewidth]{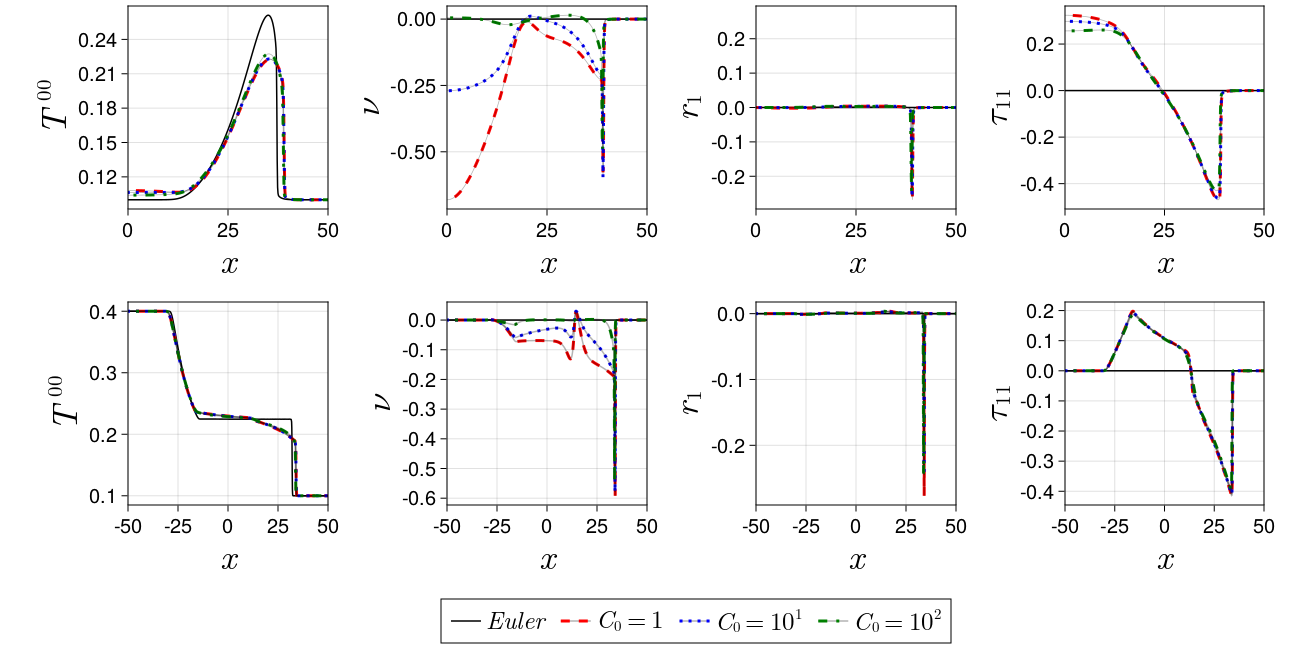}
    \caption{Energy density (gaussian profile at the top panel; discontinuous profile at the bottom panel) and the three dissipative variable profiles at $t=47\,\mbox{GeV}^{-1}$ for different values of $C_0$ and with $\chi_2 = -1$, $C_1 = 10^3$, $C_2 = 1$.}
    \label{fig:chichonstep-C2=-1-C0var-Enur1tau11}
\end{figure*}
Several effects can be noticed here. In both cases, the propagation speed of the shock gets changed, as can be seen by the change in the position of the shock. In the discontinuous initial data, a second discontinuity seems to form in $T^{00}$ for $C_2 = 0$, which seems to \textit{coincide} with a discontinuity in $\tau_{11}$. In all cases, $\tau_{11}$ greatly increases for small values of $C_2$, which suggests that this is an important variable in the evolution of the theory. This is not surprising at all, since $\tau_{ab}$ is proportional to the shear of the fluid, given that by construction it is proportional to the part of $T^{00}$ that is trace-free and perpendicular to $u^a$. The relevance of $\tau^{11}$ is numerically verified in the next subsection.
\begin{figure*}
    \centering
    \includegraphics[width =0.9\linewidth]{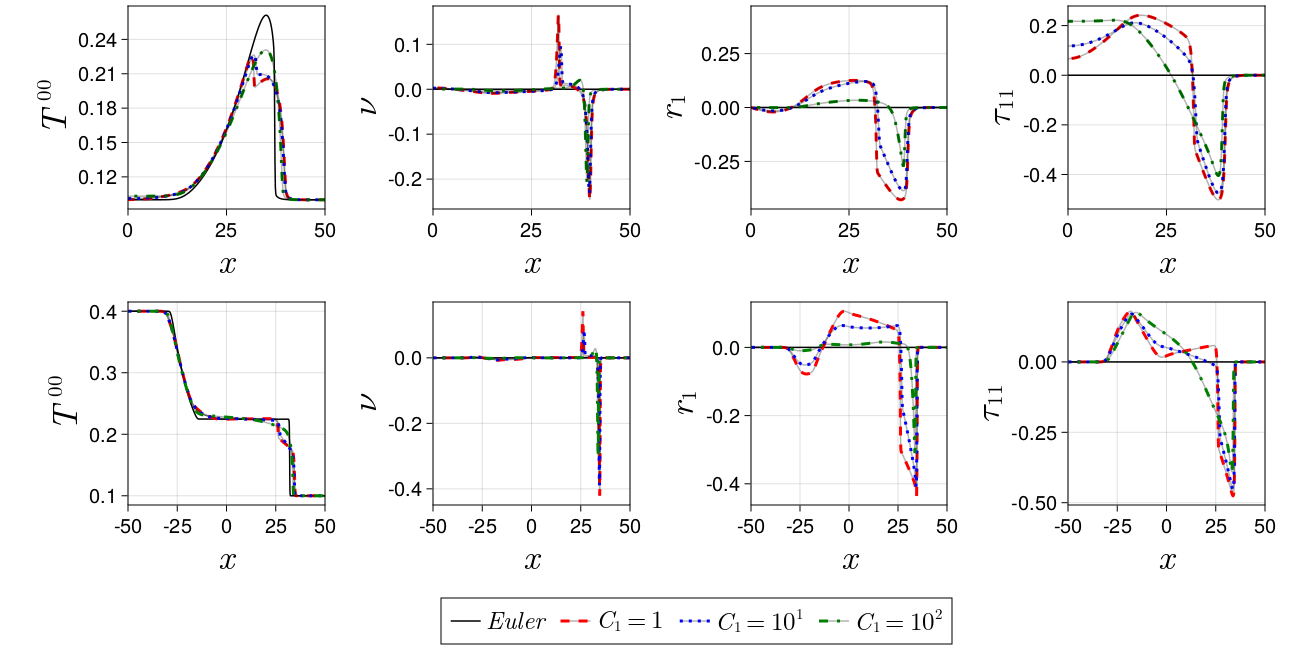}
    \caption{Energy density (gaussian profile at the top panel; discontinuous profile at the bottom panel) and the three dissipative variable profiles at $t=47\,\mbox{GeV}^{-1}$ for different values of $C_1$ and with $\chi_2 = -1$, $C_0 = 10^3$, $C_2 = 1$.}
    \label{fig:chichonstep-C2=-1-C1var-Enur1tau11}
\end{figure*}

\begin{figure*}
    \centering
    \includegraphics[width=0.9\linewidth]{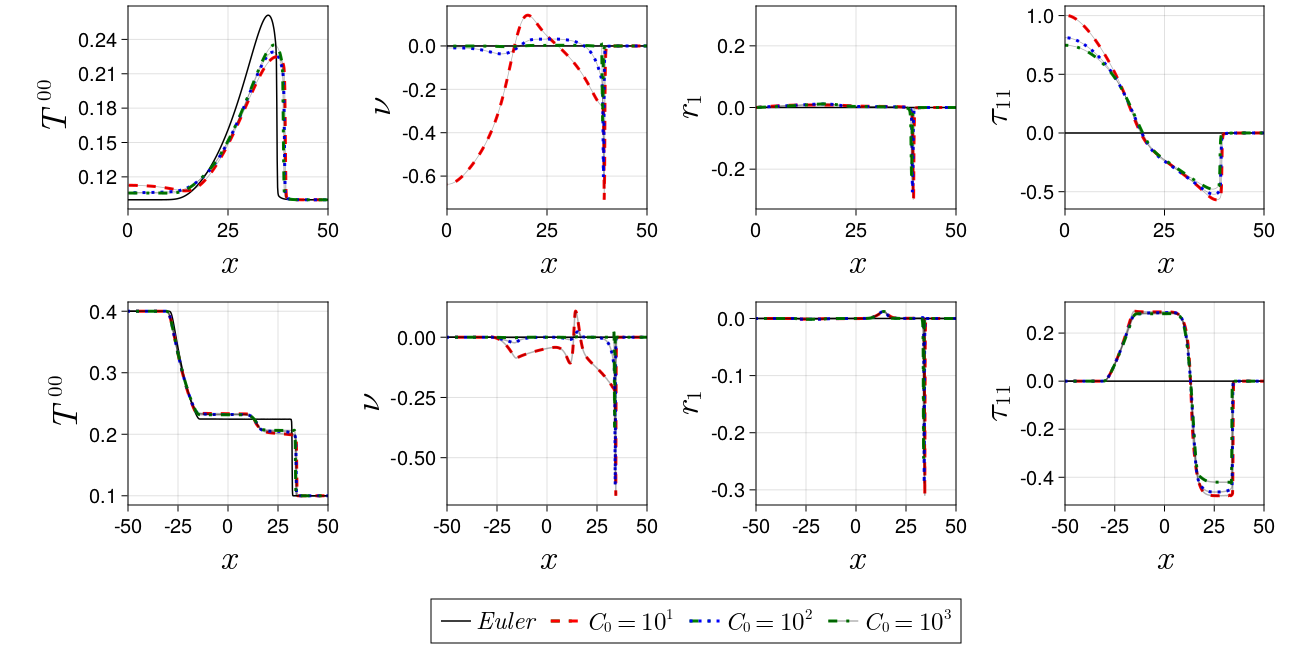}
    \caption{Energy density (gaussian profile at the top panel; discontinuous profile at the bottom panel) and the three dissipative variable profiles at $t=47\,\mbox{GeV}^{-1}$ for different values of $C_0$ and with $\chi_2 = -1$, $C_1 = 10^3$, $C_2 = 0$.}
    \label{fig:chichonstep-C2=0-C0var-Enur1tau11}
\end{figure*}

\subsubsection{What happens when varying \texorpdfstring{$C_0$}{Lg} and \texorpdfstring{$C_1$}{Lg}, but keeping \texorpdfstring{$C_2$}{Lg} small?}

As a last exploration of the parameter space, let us analyze how the system changes when keeping $C_2$ small enough, and varying the other two source parameters, given that the largest difference with respect to the perfect fluid evolution is found precisely in this regime. Figures \ref{fig:chichonstep-C2=-1-C0var-Enur1tau11} and \ref{fig:chichonstep-C2=-1-C1var-Enur1tau11} show the corresponding behaviour of the solution. Although we set $C_2 = 1$ for this analysis, we did not include here the values of $C_0$ and $C_1$ which are smaller than one, because of the presence of numerical instabilities. We notice that modifying $C_0$ results in an important change in variable $\nu$, while modifying only $C_1$ results in a very similar effect but for $r_1$. In terms of the energy density, the greatest departure from Euler´s solution can be appreciated keeping $C_1$ small enough. Both for the discontinuous and gaussian initial data we notice the formation of a second shock in $T^{00}$, which corresponds itself with a discontinuity in $\tau_{11}$.

\begin{figure*}
    \centering
    \includegraphics[width =0.9\linewidth]{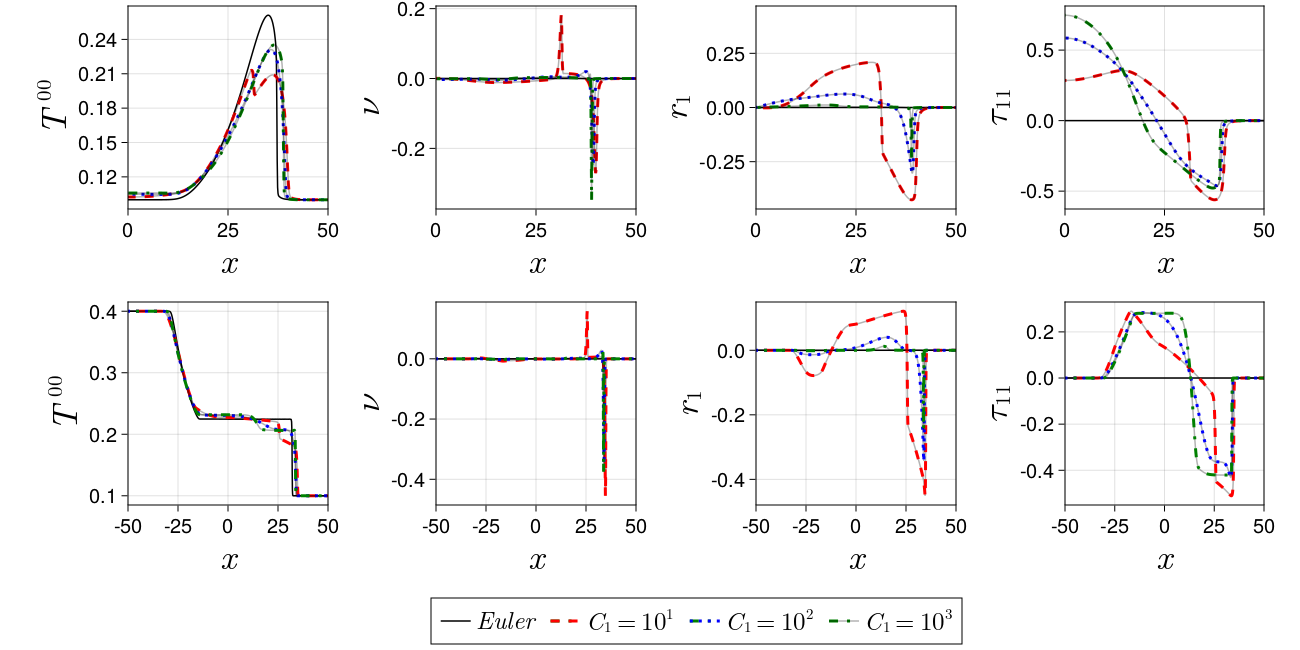}
    \caption{Energy density (gaussian profile at the top panel; discontinuous profile at the bottom panel) and the three dissipative variable profiles at $t=47\,\mbox{GeV}^{-1}$ for different values of $C_1$ and with $\chi_2 = -1$, $C_0 = 10^3$, $C_2 = 0$.}
    \label{fig:chichonstep-C2=0-C1var-Enur1tau11}
\end{figure*}

Similar results can be observed when $C_2 = 0$, as shown in figures \ref{fig:chichonstep-C2=0-C0var-Enur1tau11} and \ref{fig:chichonstep-C2=0-C1var-Enur1tau11}. In this case, numerical instabilities become present when $C_0 = 1$ and $C_1 = 1$. 

Finally, an analysis of the behaviour of the dissipative evolution variables at long times can be found in Appendix \ref{App-long-time}.

\subsection{Entropy creation rate and shock formation}

\begin{figure}
    \centering
    \includegraphics[width=\linewidth]{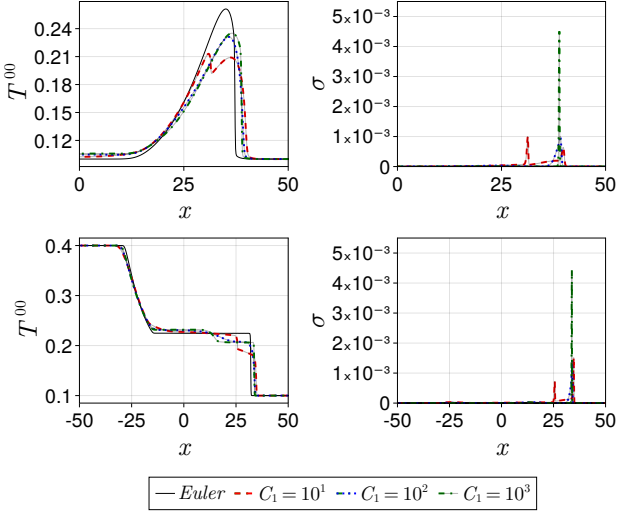}
    \caption{Entropy rate $\sigma$ at $t=47\,\mbox{GeV}^{-1}$ for different values of $C_1$, and with $C_2 = 0$, $C_0 = 10^3$.}
    \label{fig:chichonstep-C2-0-C1variable-Esigma}
\end{figure}

\begin{figure}
    \centering
    \includegraphics[width =\linewidth]{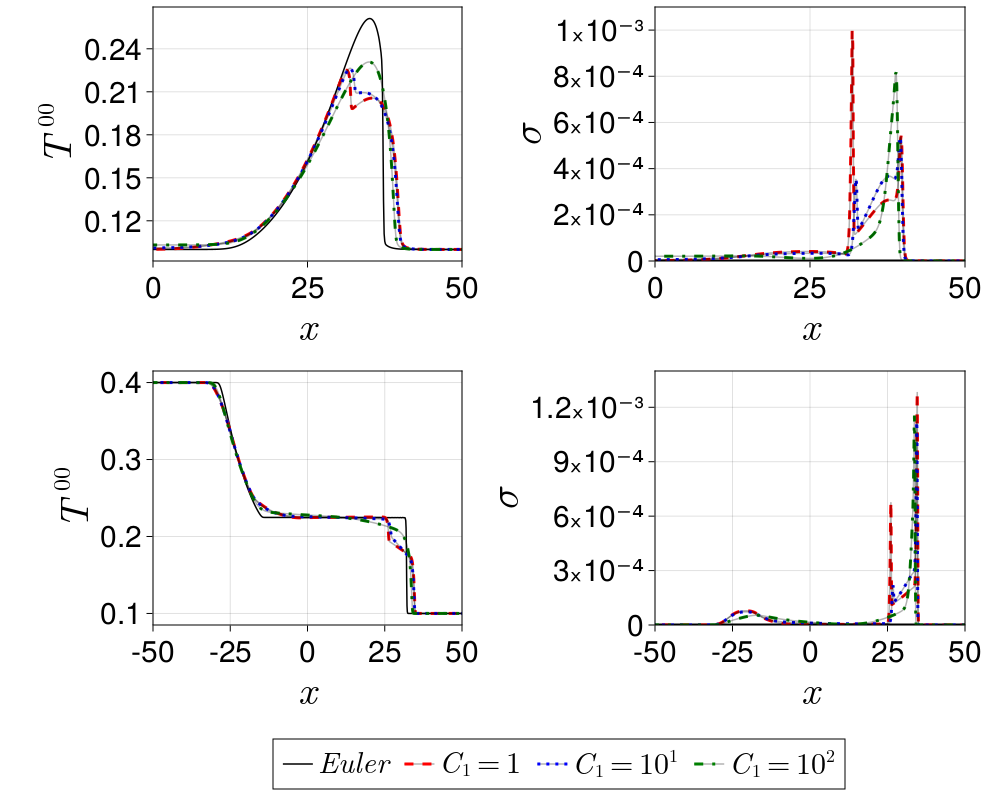}
    \caption{Entropy rate $\sigma$ at $t=47\,\mbox{GeV}^{-1}$ for different values of $C_1$ and with $C_2 = 1$, $C_0 = 10^3$.}
    \label{fig:chichonstep-C2-1-C1variable-Esigma}
\end{figure}

A very useful variable to study the formation and location of shock waves is the entropy creation rate $\sigma$, introduced in equation (\ref{entropycons}), and explicitly computed from equation (\ref{sigma-explicit}). This remarkable fact is well understood from the theory of shock formation (see \cite{Lax73,leveque92} for details). Physically, when a shock forms, the entropy of the system is expected to grow. Thus, if starting from a zero entropy-rate configuration, the formation of the shock should coincide locally with the moment in which the entropy rate changes. This quantity can be easily calculated from the dissipative variables, and has a strong correlation with the position of the peaks. As an example, figures \ref{fig:chichonstep-C2-0-C1variable-Esigma} and \ref{fig:chichonstep-C2-1-C1variable-Esigma} show a comparison between $\sigma$ and $T^{00}$ for small values of $C_2$ and for various values of $C_1$. It can be seen that the $\sigma$ distribution presents peaks in the position of the shocks in the energy distribution. This can be useful to track the position and velocity of shocks, and could be exploited by numerical schemes that need this information.


\subsection{Convergence analysis}

Finally, we study the convergence of the methods we implemented in order to evolve our fluid equations. To do so, we set $\chi_2=-1$ and $C_0=C_1=C_2=10$ for the two initial data considered throughout this numerical exploration, and perform runs for three different resolutions, taking $N=1600$, $N=3200$ and $N=6400$. Assuming that the numerical scheme is of order $p$, that is, 
\begin{equation}
||(u_N)^{k}_{j} - u_{\text{\scriptsize{exa}}}(k\Delta t ,j \Delta x)|| = \mathcal{O}(\Delta x^p),
\end{equation}
\noindent then it can be easily deduced that the convergence factor
\begin{equation}
Q(t)=\dfrac{||(u_{2N})^{k}_{j} - (u_N)^{k}_{j}||}{||(u_{4N})^{k}_{j} - (u_{2N})^{k}_{j}||} \approx 2^p.
\end{equation}
This allows us to calculate $p$ without the need of having an exact solution. To correctly study the convergence, we also need to choose an adequate norm. We chose the norm $L_{1}$, which seems to be the most appropriate for the WENO schemes, as the one we implemented here \cite{doi:10.1137/070679065}. 

\begin{figure}
\label{fig:convergence}
\includegraphics[width =\linewidth]{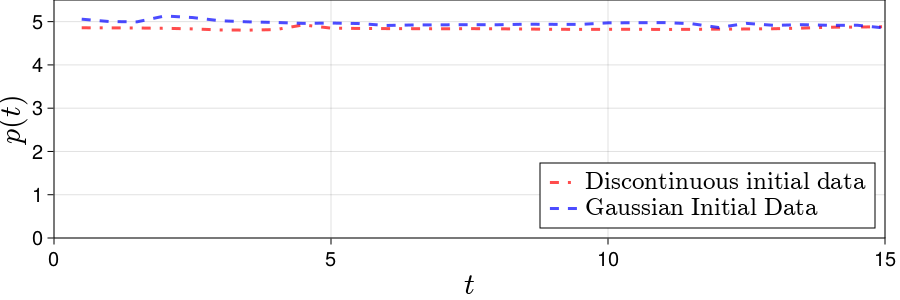}
 \caption{Convergence order $p$ as a function of evolution time. The blue dotted-line corresponds to the gaussian initial profile, while the red one is for the discontinuous data.}
\end{figure}

Results of the convergence study for $T^{00}$ can be seen in Figure \ref{fig:convergence}. Similar results can be seen for all five evolution variables, and indicate that the convergence order of the method is $p\sim 5$. Thus, our code is retaining the convergence of the WENO-Z scheme despite the need of the implementation of a Newton-Raphson scheme to calculate the fluid variables on each step.

\section{CONCLUSIONS}
\label{sec-5}

In this work we numerically evolved the divergence-type system of equations for a conformally invariant viscous relativistic fluid with zero chemical potential. We considered generating functions up to quadratic terms in the dissipative variables. This family of dissipative theories was studied in \cite{Rubio18}, where it was found that its principal part essentially depends on only one parameter. Among these theories, there is a subset with admits a well-posed initial value formulation, whose dynamical equations could serve for instance as a simple model for transport phenomena in Micro-physics.

The free parameters of the theory come from either the different contributions on the generating function ($\chi_0,\chi_1,\chi_2$), or from the source driving the dissipative variables ($C_0,C_1,C_2$). After setting $\chi_0=-1$ and $\chi_1=1$, we explored the parameter space of $\chi_2$ and $C_i$ from two different initial data sets for the energy density: (i) a smooth Gaussian function and (ii) a discontinuous function at the center of the spatial domain. We found that for large values of $\chi_2$ and $C_i$ the evolution of the system settles down to that of the Euler equations for pure radiation. This is due to the fact that, under these conditions, the magnitude of the dissipative variables becomes negligible, having almost no effect on the dynamics of the system. On the other hand, for small values of $\chi_2$ and $C_i$, the propagation speeds get notoriously changed with respect to the Euler's structure, which causes the formation of different shock waves and new peaks that are not present in the perfect-fluid dynamics. This can be understood from the fact that, unlike the ideal fluid case in which there is only one speed for the propagation of perturbations, the dissipative fluid system admits not one but two sound speeds, in addition to a standing mode.

To study the effect of the source function parameters, we fixed $\chi_2 = -1$ (as to depart from the dynamics of a pure perfect fluid), and found that the most relevant contribution was the one driven by the parameter $C_2$. In particular, when $C_2$ is large in magnitude, $C_0$ and $C_1$ have no appreciable effect on the dynamics of the solution. Notoriously, the dissipative variable whose behaviour is most affected by $C_2$ is in fact $\tau_{11}$, which within certain limit it corresponds to the shear of fluid, being $\tau_{ab}$ traceless and completely orthogonal to $u^{a}$. This behaviour seems to be characteristic when evolving a conformal theory, as it also played a relevant role in the dynamics of other types of conformal theories \cite{Pretorius21,Pretorius22,Figueras22}.
After that, by fixing $C_2=0$ and $C_2=1$ we reported new effects when varying $C_0$ and $C_1$, particularly in the later case, where we even observed the formation of new shocks during the evolution, presumably corresponding to the extra degrees of freedom of the theory.

Our system also allows for an easy calculation of the entropy generation rate $\sigma$, which displays peaks in the shock regions. This can be very useful for keeping track on shock formation.

 The experience gained in this work suggests the possibility of generalizing this scheme to full 3D simulations, accounting also for curved spacetimes or even fully relativistic (astrophysical/cosmological) scenarios. Since the extension from the present work to a curved background should be straightforward, we plan to use this theory as a matter source for solving the full case considering backreaction with the spacetime geometry, as well as including baryon current density, i.e., studying dissipative fluids in the context of strong gravity (as a source for modeling neutron stars, black hole accretion disks, among other applications).

Finally, it would be interesting to study the plausibility of a ``reduced'' theory; i.e., where only shear-like degrees of freedom are exited, but no heat-flux ones. Unlike the theory simulated in this paper (in which more degrees of freedom are always exited out of generic initial data), it might be that for the particular value of the parameter $\chi_2$ at which the transformation between conservative and abstract variables breaks down, there is a smaller set of degrees of freedom that could get exited in a generic way. This is part of ongoing work.

\section*{Acknowledgements}

We thank Fernando Abalos, Miguel Bezares and Federico Carrasco for a careful reading of a first version of this manuscript, and for their valuable comments and suggestions. PM and OR acknowledge financial support from CONICET, SeCyT-UNC, and MinCyT-Argentina. MR acknowledges support from the European Union’s H2020 ERC Consolidator Grant ``GRavity from Astrophysical to Microscopic Scales'' (Grant No. GRAMS-815673) and the EU Horizon 2020 Research and Innovation Programme under the Marie Sklodowska-Curie Grant Agreement No. 101007855. 

\appendix

\section{Explicit form of the evolution equations}
\label{app-expeqs}

In this appendix we give the explicit expressions for the relevant components of the tensors $T^{ab}$ and $A^{abc}$ that were used for both the inversion of the transformation between the fluid variables $\{\mu,v,\nu,r^1,\tau^{11}\}$ and the conservative ones $\{T^{00},T^{01},A^{000},A^{001},A^{011}\}$, and the non trivial flux functions $\{T^{11},A^{111}\}$ in the evolution equations (\ref{1+1-evo-eqs}) simulated in this work. As before, we generically write the tensor fields $T^{ab}$ and $A^{abc}$ as
\begin{eqnarray*}
    T^{ab} &=& T_0^{ab} + T_1^{ab} + T_2^{ab}\,,\\
    A^{abc} &=& A_1^{abc} + A_2^{abc},
\end{eqnarray*}
where the sub-indices stand, respectively, for the zeroth, first and second order contributions from the generating function given in eq. (\ref{chiexpansion}).

\subsection{Transformation of variables}

The explicit relation between the conservative and fluid variables is given by the following formulae:
\begin{alignat*}{1}
    T^{00} &= -2 \chi _0 \left(4 \gamma ^2-1\right) T^4 \\ 
    & -\frac{2}{3} T^6 \chi _1 \left(30 \gamma  r_1 T v+10 \left(4 \gamma ^2-1\right) \nu  T^2+3 \tau _{11} v^2\right) \\
    & -\frac{1}{3} T^8 \chi _2 \left(336 \gamma  r_1 T v \left(10 \nu  T^2-3 \tau _{11} \left(v^2-1\right)\right)\right.\\ 
    &\left. -36 r_1^2 T^2 \left(8 \gamma ^4-51 \gamma ^2+\gamma ^2
   \left(8 \gamma ^2-3\right) v^4\right.\right. \\
   & \left.\left. -2 \left(8 \gamma ^4-27 \gamma ^2+12\right) v^2+8\right)\right.\\
   & \left. +16 \nu ^2 T^4 \left(16 \gamma ^6 \left(v^2-1\right)^2+\gamma ^2
   \left(243-19 v^2\right)\right.\right. \\
   &\left.\left. -2 \gamma ^4 \left(v^4-52 v^2+51\right)-52\right)+336 \nu  \tau _{11} T^2 v^2\right. \\
   & \left. +9 \tau _{11}^2 \left(v^2-1\right) \left(-24 \gamma
   ^2+\left(24 \gamma ^2-11\right) v^2+3\right)\right)\,;
\end{alignat*}
\begin{alignat*}{1}
    T^{01} &= -8 \chi _0\gamma ^2 T^4 v \\ 
    & + \chi _1 T^6\left(-10 \gamma  r_1 T \left(v^2+1\right)-\frac{80}{3} \gamma ^2 \nu  T^2 v-2 \tau _{11} v\right) \\
    & -\chi _2\frac{8}{3} T^8 \left(3 \gamma  r_1 T \left(v^2+1\right) \left(10 \nu  T^2 \left(2 \gamma ^2 \left(v^2-1\right)+9\right)\right.\right.\\
    &\left.\left. -21 \tau _{11}
   \left(v^2-1\right)\right)-36 r_1^2 T^2 v \left(\gamma ^4 \left(v^2-1\right)^2+6 \gamma ^2 \left(v^2-1\right) \right.\right.\\
   &\left.\left. -2\right)+v \left(8 \gamma ^2 \nu ^2 T^4 \left(4
   \gamma ^4 \left(v^2-1\right)^2+25 \gamma ^2 \left(v^2-1\right)+56\right) \right.\right.\\
   & \left.\left. +42 \nu  \tau _{11} T^2+9 \tau _{11}^2 \left(v^2-1\right) \left(3 \gamma ^2
   \left(v^2-1\right)-1\right)\right)\right)\,;
\end{alignat*}
\begin{alignat*}{1}
    A^{000} &= 3 \chi _1 \gamma  \left(1-2 \gamma ^2\right) T^5 \\ 
    & -12 T^8 \chi _2 \left(3 \left(6 \gamma ^2-1\right) r_1 v+10 \gamma  \left(2 \gamma ^2-1\right) \nu  T+\frac{3 \gamma  \tau _{11} v^2}{T}\right)\,;
\end{alignat*}
\begin{alignat*}{1}
    A^{001} &= \gamma  \left(1-6 \gamma ^2\right) T^5 v \chi _1 \\ 
    & -12 T^8 \chi _2 \left(r_1 \left(6 \gamma ^2 \left(2 v^2+1\right)-1\right)+\frac{\gamma  \tau _{11} v \left(v^2+2\right)}{T} \right. \\
    & \left. +\frac{10}{3} \gamma  \left(6 \gamma
   ^2-1\right) \nu  T v\right)\,;
\end{alignat*}
\begin{alignat*}{1}
    A^{011} &= -\chi_1\gamma T^5 \left(6 \gamma ^2 v^2+1\right) \\ 
    & -12 T^8 \chi _2 \left(r_1 v \left(6 \gamma ^2 \left(v^2+2\right)+1\right)\right.\\
    &\left. +\frac{10}{3} \gamma  \nu  T \left(6 \gamma ^2 v^2+1\right) +\frac{\gamma  \tau _{11} \left(2
   v^2+1\right)}{T}\right).
\end{alignat*}

\subsection{Explicit formulae for \texorpdfstring{$T^{11}$}{Lg} and \texorpdfstring{$A^{111}$}{Lg}}

At each time step, and after numerically inverting the relations shown in the previous subsection, we evaluated the non trivial numerical fluxes by using the following expressions:
\begin{alignat*}{1}
    T^{11} &= -2 \chi _0 T^4 \left(4 \gamma ^2 v^2+1\right) \\ 
    & -\frac{2}{3} T^6 \chi _1 \left(30 \gamma  r_1 T v+3 \tau _{11}+10 \nu  T^2 \left(4 \gamma ^2 v^2+1\right)\right) \\
    & -\frac{1}{3} T^8 \chi _2 \left(336 \gamma  r_1 T v \left(10 \nu  T^2-3 \tau _{11} \left(v^2-1\right)\right)\right. \\
    & \left. -36 r_1^2 T^2 \left(3 \left(\gamma ^2-8\right)+8 \gamma ^4
   v^6+\left(51 \gamma ^2-16 \gamma ^4\right) v^4 \right.\right.\\
   & \left.\left. +\left(8 \gamma ^4-54 \gamma ^2+8\right) v^2\right)+16 \nu ^2 T^4 \left(16 \gamma ^6 v^2 \left(v^2-1\right)^2\right.\right.\\
   & \left.\left. +\gamma
   ^2 \left(243 v^2-19\right)+2 \gamma ^4 \left(51 v^4-52 v^2+1\right)+52\right)\right.\\
   & \left. +336 \nu  \tau _{11} T^2+9 \tau _{11}^2 \left(v^2-1\right) \left(24 \gamma ^2
   v^4+\left(3-24 \gamma ^2\right) v^2 \right.\right.\\
   &\left.\left.-11\right)\right)\,;
\end{alignat*}
\begin{alignat*}{1}
    A^{111} &= -3 \gamma  T^5 v \chi _1 \left(2 \gamma ^2 v^2+1\right) \\ 
    & -12 T^8 \chi _2 \left(3 r_1 \left(6 \gamma ^2 v^2+1\right)+10 \gamma  \nu  T v \left(2 \gamma ^2 v^2+1\right) \right. \\
    & \left. +\frac{3 \gamma  \tau _{11} v}{T}\right).
\end{alignat*}

\section{WENO-Z method}
\label{app-wenoz}
We give a brief review of Weighted Essentially Non-Oscillatory (WENO) schemes, particularizing to the WENO-Z scheme. A clear and detailed discussion on this family of high-order schemes can be found in the review \cite{SHU1988439}, as well as some applications of WENO schemes for the evolution of binary systems were performed in \cite{Palenzuela18,Bernuzzi16,Liebling2020}.

To evolve the system of equations studied throughout this work, we implemented the WENO-Z method, introduced by Borges et. al. in \cite{BORGES20083191}. This method keeps track to more general WENO scheems for the numerical integration of systems of conservation laws, which are of the form (in 1D, for simplicity)
\begin{equation}
\partial_t u + \partial_x F(u(x)) = g(u(x)).
\label{eq:conservation-law}
\end{equation}

We discretize the above equation by using an evenly spaced grid $x_{j}:=j\Delta x$. As these systems generically develop shocks, we cannot use straightforward finite differences to directly calculate $\partial_x F(u)$. This suggests to consider the semi-discreet approximation given by
\begin{equation}
	\dfrac{du_j}{d t} + \dfrac{\hat F_{j+1/2} - \hat F_{j-1/2}}{\Delta x} = g(u(x_j)),
\end{equation}
where $u_j(t):= u(x_j,t)$. The main issue here is how to define (or reconstruct) the numerical discretization of the flux function $F(u)$, i.e., how to find $\hat F_{j+1/2}$. By implicitly defining a function $h(x)$ such that
\begin{equation}
	\dfrac{1}{\Delta x}\int_{x-\Delta x/2}^{x+\Delta x/2}h(\xi)d\xi = F(u(x)),
\end{equation}
we get then that 
\begin{equation}
   \partial_x F = h(x_{j+1/2})-h(x_{j-1/2}),
\end{equation}
which implies that 
\begin{equation}
    \hat{F}_{j+1/2} = h(x_{j+1/2}).   
\end{equation}
Notice then that $F(u(x_j))$ corresponds to the \textit{cell average} of the function $h$. Thus, what one needs at the end is a way to approximate the function $h$ in terms of $F(u(x_j))$. It is exactly for this step that WENO schemes are useful for. Indeed, it is possible to reconstruct a quantity $v_{j+1/2}$ from the cell averages $\bar{v}_{j+k}$ ($k=-2,-1,0,1,2$); i.e., by using a stencil of five points. Firstly, a third-order reconstruction of $v_{j+1/2}$ using only three cell averages in three different stencils is performed, namely
\begin{eqnarray}\label{eq-disc}
    v^{(1)}_{j+1/2} &=& \dfrac{1}{3}\bar{v}_{j-2} -\dfrac{7}{6}\bar{v}_{j-1} +  \dfrac{11}{6}\bar{v}_{j},\\
    v^{(2)}_{j+1/2} &=& -\dfrac{1}{6}\bar{v}_{j-1} +\dfrac{5}{6}\bar{v}_{j} +  \dfrac{1}{3}\bar{v}_{j+1},\\
    v^{(3)}_{j+1/2} &=& \dfrac{1}{3}\bar{v}_{j} +\dfrac{5}{6}\bar{v}_{j+1} -  \dfrac{1}{6}\bar{v}_{j+2}.
\end{eqnarray}
Then, it can be shown that, if $v$ is sufficiently smooth, $v^{(k)}_{j+1/2} = v(x_{j+1/2}) + \mathcal{O}(\Delta x^3)$. Now, if there happens to be a discontinuity in the physical solution, the scheme chooses the stencil where the discontinuity is \textit{not} present, and normally continue with the integration. The way WENO schemes achieve this is by weighting each approximation and then adding all them up, getting
\begin{equation}
v_{j+1/2} = \sum_{k=1}^{3}w_{k}v^{(k)}_{j+1/2},
\label{eq:weno-sum}
\end{equation}
where $\sum w_{k} = 1$ and $w_k$ are smooth functions of $v_{j-2}$, $v_{j-1}$, $v_{j}$, $v_{j+1}$ and $v_{j+2}$, in such a way that, if $v$ is smooth, the linear combination \eqref{eq:weno-sum} is $\mathcal{O}(\Delta x^5)$; while if a discontinuity is present in a particular cell, the stencil (or stencils) where such discontinuity is present has $w^{(k)} = 0$, becoming $v_{j+1/2} = v(x_{j+1/2})+\mathcal{O}(\Delta x^3)$. 

Different WENO reconstruction schemes differ among them in the way these weights are actually computed. The WENO-Z scheme, for instance, works as follows:

\begin{itemize}
\item[1.] Compute the following three \textit{smoothness} indicators:
\begin{eqnarray*}
	\beta^{(1)} &=& \dfrac{13}{12}(v_{i-2}-2v_{i-1}+v_{i})^2+\dfrac{1}{4}(v_{i-2}-4v_{i-1}+3v_{i})^2 \\
    \beta^{(2)} &=& \dfrac{13}{12}(v_{i-1}-2v_{i}+v_{i+1})^2+\dfrac{1}{4}(v_{i-1}-v_{i+1})^2 \\
    \beta^{(3)} &=& \dfrac{13}{12}(v_{i}-2v_{i+1}+v_{i+2})^2+\dfrac{1}{4}(3v_{i}-4v_{i+1}+v_{i+2})^2.
\end{eqnarray*}

\item[2.] Calculate the weights from the following formula:
\begin{equation}
	w_{k} = \dfrac{\alpha_{k}}{\sum_{l=1}^{3}\alpha_{l}},\quad \alpha_k = d_{k}\left( 1+\dfrac{\tau_{5}}{\beta_{k}+\epsilon}\right),\quad k=1,2,3
\end{equation}
where $\tau_5 = |\beta_{1}-\beta_{3}|$, $\epsilon$ is a small number in order to avoid divisions by zero, and $d_{k}$ the coefficients for which $\sum_{k=1}^{3}d_{k}v_{j+1/2}^{(k)} = v(x_{j+1/2})+\mathcal{O}(\Delta x^5)$,
\begin{equation}
d_1 = \dfrac{1}{16}, \quad d_2 = \dfrac{5}{8},\quad d_3 = \dfrac{5}{8}.
\end{equation}
For our simulations, we set $\epsilon = 10^{-40}$, although it can be easily checked that the results do not get considerably affected.

\item[3.] Compute $\hat F_{j+1/2} = h(x_{j+1/2})$. 
\end{itemize}

Notice that since that there is one more point to the left of $j+1/2$, the scheme is ``biased'' to the left. Of course one can also create a ``right-biased'' scheme, just by switching the order of the arguments, namely,
\begin{align*}
\hat{F}_{j+1/2}(u_{j-2}, u_{j-1}, u_{j}, u_{j+1}, u_{j+2}) \quad\\ \rightarrow \hat{F}_{j+1/2}(u_{j+2}, u_{j+1}, u_{j}, u_{j-1}, u_{j-2}).
\end{align*}

Finally, once the flux was computed at a time step, we proceed with the time integration, using the optimal 3-stage, third order SSP Runge-Kutta method (SSPRK33) \cite{SHU1988439}. By introducing the operator
\begin{equation}
    L(u_i(t)):= - \frac{\hat{F}_{i+\frac{1}{2}} - \hat{F}_{i-\frac{1}{2}}}{\Delta x} + g(u_i(t)),
\end{equation}
from equation (\ref{eq-disc}), the RK steps are
\begin{align*}
    u_{i}^{(1)} &= u_{i}^{n} + \Delta t L(u^{n}),\\
    u_{i}^{(2)} &= \dfrac{3}{4}u_{i}^{n} + \dfrac{1}{4}u_{i}^{(1)} + \dfrac{1}{4} \Delta t L(u^{(1)}),\\
    u_{i}^{n+1} &= \dfrac{1}{3}u_{i}^{n}+ \dfrac{2}{3}u_{i}^{(2)} + \dfrac{2}{3}\Delta t L(u^{(2)}).
\end{align*}
This particular method allowed us to evolve the conservative variables while capturing shock propagation, preserving the TVD property of high-order schemes for system of conservation laws.

\section{A curiosity on the transformation from fluid to conservative variables}\label{cuenta-pablo}

In this appendix we show, by means of a simple construction, that the Jacobian of the transformation between fluid and conservative variables used throughout this work is not bijective as a function of the dissipative parameter $\chi_2$. This surprising fact came out while exploring the propagation speeds for different values of such parameter. Here we provide a procedure to find its value. 

Firstly, a direct inspection of the formulas for $A^{abc}$ and $T^{ab}$ allows to notice the following relations (which hold off-shell; that is, without using the dynamical equations):
\begin{eqnarray}
    A_{1}^{abc} \xi_{c} &=& -\dfrac{\chi_{1}}{2\chi_{0}}T_{0}^{ab}\\
    A_{2}^{abc}\xi_{c} &=& -\dfrac{6\chi_{2}}{\chi_{1}}T_{1}^{ab}
    \label{eq:A_T_comparisson}
\end{eqnarray}

Then, we focus on a simple model for the transformation function between the conservative variables  $\{T^{00}, T^{01}, A^{000}, A^{001},A^{011}\}$, and the abstract variables, $\{\xi^0, \xi^1, \xi^{00}, \xi^{01}, \xi^{11}\}$. In particular, we propose the simplest possible relation, which takes the form
\begin{eqnarray} \label{A-proposal}
    A_1 &=& f(x)\\
    A_2 &=& g(x)y,    
\end{eqnarray}
where $x$ represents the quantity $\xi_{a}$ and $y$ represents the dissipative variables $\xi_{ab}$. This combined with \eqref{eq:A_T_comparisson} indicates that we can write $T$ in terms of $f$ and $g$, namely
\begin{eqnarray}
    T_0 &=& -\dfrac{2\chi_{0}}{\chi_{1}}xf(x)\\
    T_1 &=& -\dfrac{\chi_{1}}{6\chi_{2}}xg(x)y.
\end{eqnarray}
Therefore (and suppressing tensorial indices for shortness of notation), at second order in $y$ we get
\begin{eqnarray}
A &=& f(x)+g(x)y\\
T &=& \alpha xf(x) + \beta xg(x)y + \mathcal{O}(y^{2}),
\end{eqnarray}
where the scaling coefficients $\alpha$ and $\beta$ are given by
\begin{equation}
    \alpha = -\dfrac{2\chi_{0}}{\chi_{1}}, \;\; \beta = -\dfrac{\chi_{1}}{6\chi_{2}}.
\end{equation}
Since we want to derive with respect to $x$ and $y$ and then evaluate at $y=0$ (that is, we are just looking for the critical $\chi_2$ value when $\xi_{ab} = 0$), we can get rid of higher-order contributions in $y$.

The Jacobian of this transformation at equilibrium (i.e., at $y = 0$), reads
\begin{equation}
    \mathbb{J} = \begin{bmatrix}
 \alpha(f(x)+xf'(x))& \beta xg(x)\\ 
 f'(x)&g(x) 
\end{bmatrix}.
\end{equation}
One can easily see that the rows of this matrix become linearly \textit{dependent} if and only if 
\begin{equation}\label{cond-no-bije}
    \alpha\dfrac{f(x)+xf'(x)}{f'(x)}=\beta x,
\end{equation}
and that will be the case when the transformation of variables is \textit{not} bijective. From eqs. (\ref{eq:A_T_comparisson}) and (\ref{A-proposal}), we find that $f(x)$ must be of the form 
\begin{equation}
    f(x) = Kx^{-5},
\end{equation}
for some real constant $K$, and therefore $xf'(x) = -5f(x)$. Replacing this in eq. (\ref{cond-no-bije}) gives the conditions
\begin{eqnarray}
\alpha \dfrac{4}{5} &=& \beta\\
\dfrac{\chi_{2}\chi_{0}}{\chi_{1}^{2}}&=& \frac{5}{48} 
\label{eq:548-condition}
\end{eqnarray}
In particular, choosing $\chi_0=-1$ and $\chi_1=1$, we get $\chi_2=-5/48$. This means that we cannot choose the parameters where condition (\ref{eq:548-condition}) is satisfied. This fact seems to suggest that, at least close to equilibrium, the system can be described by less variables, and correspondingly less equations. The study of this property is part of a work in progress.

\section{Long-time evolution for the conserved quantities}
\label{App-long-time}

In this appendix we show how is the behaviour of the conserved variables defined from the dynamical fields considered for the evolution. In particular, we stress out that, even if the dissipative variables are initially set to zero, they can settle into a (non-zero) constant value at long times, showing that the dissipative effect is not a transient effect, but it keeps during the evolution, as a consequence of the dynamical equations. 

Recall that if $u(t,x)$ obeys an equation of the form
\begin{equation}
    \partial_t u + \partial_x f(u) = g(u)
\end{equation}
then the quantity
\begin{equation}
    U(t)=\int_D{u(t,x)\,dx}
\end{equation}
satisfies the following conservation law
\begin{equation}
    \frac{dU}{dt}-\int_D{g(u(t,x))dx}=0,
\end{equation}
where $D$ is the spacial domain on which $u$ is defined. In particular, if $g=0$, $U$ is a constant of motion. We study the long-time behaviour of such conserved quantities (that is, the integral of the conservative variables of the theory), and verify that, in particular, the ones coming from the dissipative variables do not decay to zero, and moreover, they have an exponentially tendency, as expected.

\begin{figure}
    \centering
    \includegraphics[width = 0.9\linewidth]{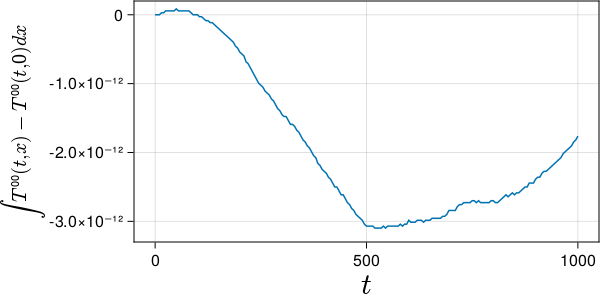}
    \caption{Integral in space of the difference $T^{00}(t,x)-T^{00}(t=0,x)$, from $t = 0$ to $t=1000$.}
    \label{fig:conservation-long-1}
\end{figure}

During the evolution, the magnitude of the conservative quantities $A^{000}$ and $A^{001}$ decreases in absolute value. However, this does not mean that they go to zero. As an example, let us study the evolution of the gaussian peak for long times, with parameters $\chi_2 = -1$ and $C_0 = 1000$, $C_1=1000$ and $C_2 = 1$. Since in the evolution equations both $T^{00}$ and $T^{01}$ do not have a source, their integrals are constant over time. Besides, both the integrals of $T^{01}$ and $A^{001}$ are trivially zero since they are odd functions in space. Figure \ref{fig:conservation-long-1} shows the integral (in space) of the difference between $T^{00}(t,x)$ and its initial value, namely $T^{00}(t=0,x)$, with $t$ ranging from 0 to 1000. We can see that it initially decreases slightly and later starts to grow, which we attribute to numerical errors (in particular, the one coming from the Newton-Raphson inversion scheme).

\begin{figure}
    \centering
    \includegraphics[width = \linewidth]{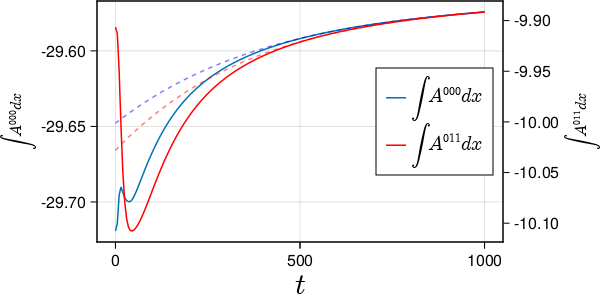}
    \caption{Integral in space of the dissipative evolution fields $A^{000}$ and $A^{011}$ over time (continuous lines) and exponential fits (dashed lines) for long times.}
    \label{fig:conservation-long-2}.
\end{figure}

On the other hand, the integrals of $A^{000}$ and $A^{011}$ are not constant but after some time, they settle exponentially to a constant, as shown in figure \ref{fig:conservation-long-2}.

Thus, the fact that the dissipative variables are initially zero does not mean that $A^{abc} = 0$, since $A^{abc}$ does not go to zero when $\xi^{ab} = 0$. That is also the reason why for long times $A^{abc}$ does not go to zero but to a constant stationary value.

\bibliographystyle{ieeetr}
\bibliography{biblio}

\begin{thebibliography}{10}

\bibitem{Rubio18}
L.~Lehner, O.~A. Reula, and M.~E. Rubio, ``{Hyperbolic theory of relativistic
  conformal dissipative fluids},'' {\em Phys. Rev.}, vol.~D97, no.~2,
  p.~024013, 2018.

\bibitem{Netzer13}
H.~Netzer, {\em The Physics and Evolution of Active Galactic Nuclei}.
\newblock Cambridge University Press, 2013.

\bibitem{Blandford19}
R.~Blandford, D.~Meier, and A.~Readhead, ``Relativistic jets from active
  galactic nuclei,'' {\em Annual Review of Astronomy and Astrophysics},
  vol.~57, no.~1, pp.~467--509, 2019.

\bibitem{Charlet21}
A.~Charlet, R.~Walder, A.~Marcowith, D.~Folini, J.~M. Favre, and M.~E.
  Dieckmann, ``{Effects of radiative losses on the relativistic jets of
  high-mass microquasars},'' {\em Astron. Astrophys.}, vol.~658, p.~A100, 2022.

\bibitem{Meier03}
D.~L. Meier, ``The theory and simulation of relativistic jet formation: towards
  a unified model for micro- and macroquasars,'' {\em New Astronomy Reviews},
  vol.~47, no.~6, pp.~667 -- 672, 2003.
\newblock The physics of relativistic jets in the CHANDRA and XMM era.

\bibitem{Peer15}
A.~Pe'er, H.~Barlow, S.~O'Mahony, R.~Margutti, F.~Ryde, J.~Larsson, D.~Lazzati,
  and M.~Livio, ``{HYDRODYNAMIC} {PROPERTIES} {OF} {GAMMA}-{RAY} {BURST}
  {OUTFLOWS} {DEDUCED} {FROM} {THE} {THERMAL} {COMPONENT},'' {\em The
  Astrophysical Journal}, vol.~813, p.~127, nov 2015.

\bibitem{Ramirez10}
E.~Ramirez-Ruiz and A.~I. MacFadyen, ``{THE} {HYDRODYNAMICS} {OF} {GAMMA}-{RAY}
  {BURST} {REMNANTS},'' {\em The Astrophysical Journal}, vol.~716,
  pp.~1028--1039, may 2010.

\bibitem{Raposo19}
G.~Raposo, P.~Pani, M.~Bezares, C.~Palenzuela, and V.~Cardoso, ``Anisotropic
  stars as ultracompact objects in general relativity,'' {\em Phys. Rev. D},
  vol.~99, p.~104072, May 2019.

\bibitem{Stone01}
J.~M. Stone and J.~E. Pringle, ``{Magnetohydrodynamical non-radiative accretion
  flows in two dimensions},'' {\em Monthly Notices of the Royal Astronomical
  Society}, vol.~322, pp.~461--472, 04 2001.

\bibitem{Yuan14}
F.~Yuan and R.~Narayan, ``Hot accretion flows around black holes,'' {\em Annual
  Review of Astronomy and Astrophysics}, vol.~52, no.~1, pp.~529--588, 2014.

\bibitem{Das07}
S.~Das, ``{Behaviour of dissipative accretion flows around black holes},'' {\em
  Monthly Notices of the Royal Astronomical Society}, vol.~376, pp.~1659--1670,
  04 2007.

\bibitem{Giustini19}
M.~Giustini and D.~Proga, ``A global view of the inner accretion and ejection
  flow around supermassive black holes. radiation-driven accretion disk winds
  in a physical context,'' {\em Astronomy \& Astrophysics}, vol.~630, 08 2019.

\bibitem{Abbott21}
R.~A. et. al., ``Observation of gravitational waves from two neutron
  star{\textendash}black hole coalescences,'' {\em The Astrophysical Journal
  Letters}, vol.~915, p.~L5, jun 2021.

\bibitem{GW170817}
T.~L.~S. Collaboration and V.~Collaboration, ``Gw170817: Observation of
  gravitational waves from a binary neutron star inspiral,'' {\em Phys. Rev.
  Lett.}, vol.~119, p.~161101, Oct 2017.

\bibitem{Chatziioannou18}
K.~Chatziioannou, C.-J. Haster, and A.~Zimmerman, ``{Measuring the neutron star
  tidal deformability with equation-of-state-independent relations and
  gravitational waves},'' {\em Phys. Rev. D}, vol.~97, no.~10, p.~104036, 2018.

\bibitem{Landry20}
P.~Landry, R.~Essick, and K.~Chatziioannou, ``{Nonparametric constraints on
  neutron star matter with existing and upcoming gravitational wave and pulsar
  observations},'' {\em Phys. Rev. D}, vol.~101, no.~12, p.~123007, 2020.

\bibitem{Busza18}
W.~Busza, K.~Rajagopal, and W.~van~der Schee, ``{Heavy Ion Collisions: The Big
  Picture, and the Big Questions},'' {\em Ann. Rev. Nucl. Part. Sci.}, vol.~68,
  pp.~339--376, 2018.

\bibitem{DEnterria06}
D.~G. d'Enterria, ``{High-energy heavy-ions physics: From RHIC to LHC},'' {\em
  Nucl. Phys. A}, vol.~782, pp.~215--223, 2007.

\bibitem{McDonough20}
E.~McDonough, ``{The Cosmological Heavy Ion Collider: Fast Thermalization after
  Cosmic Inflation},'' {\em Phys. Lett. B}, vol.~809, p.~135755, 2020.

\bibitem{Calzetta99}
E.~A. Calzetta, B.~L. Hu, and S.~A. Ramsey, ``{Hydrodynamic transport functions
  from quantum kinetic theory},'' {\em Phys. Rev. D}, vol.~61, p.~125013, 2000.

\bibitem{Peralta11}
J.~Peralta-Ramos and E.~Calzetta, ``{Shear viscosity from thermal fluctuations
  in relativistic conformal fluid dynamics},'' {\em JHEP}, vol.~02, p.~085,
  2012.

\bibitem{Calzetta14}
E.~Calzetta, ``{Hydrodynamic approach to boost invariant free streaming},''
  {\em Phys. Rev. D}, vol.~92, no.~4, p.~045035, 2015.

\bibitem{Elias14}
M.~Elias, J.~Peralta-Ramos, and E.~Calzetta, ``{Heavy quark collisional energy
  loss in the quark-gluon plasma including finite relaxation time},'' {\em
  Phys. Rev. D}, vol.~90, no.~1, p.~014038, 2014.

\bibitem{Ollitrault07}
J.-Y. Ollitrault, ``{Relativistic hydrodynamics for heavy-ion collisions},''
  {\em Eur. J. Phys.}, vol.~29, pp.~275--302, 2008.

\bibitem{Jaiswal16}
A.~Jaiswal and V.~Roy, ``{Relativistic hydrodynamics in heavy-ion collisions:
  general aspects and recent developments},'' {\em Adv. High Energy Phys.},
  vol.~2016, p.~9623034, 2016.

\bibitem{Calzetta199}
E.~Calzetta and L.~Cantarutti, ``{Dissipative type theories for Bjorken and
  Gubser flows},'' {\em Int. J. Mod. Phys. A}, vol.~35, no.~14, p.~2050074,
  2020.

\bibitem{Calzetta21}
E.~Calzetta, ``{Steady asymptotic equilibria in conformal relativistic
  fluids},'' {\em Phys. Rev. D}, vol.~105, no.~3, p.~036013, 2022.

\bibitem{geroch1990dissipative}
R.~Geroch and L.~Lindblom, ``Dissipative relativistic fluid theories of
  divergence type,'' {\em Physical Review D}, vol.~41, no.~6, p.~1855, 1990.

\bibitem{landau}
L.~D. Landau and E.~M. Lifshitz, ``Fluid mechanics, second edition: Volume 6
  (course of theoretical physics),'' {\em Rev. Mod. Phys.}, vol.~74,
  pp.~775--823, Jul 2002.

\bibitem{rezzolla2013relativistic}
L.~Rezzolla and O.~Zanotti, {\em Relativistic Hydrodynamics}.
\newblock OUP Oxford, 2013.

\bibitem{Font08}
J.~A. Font, ``{Numerical Hydrodynamics and Magnetohydrodynamics in General
  Relativity},'' {\em Living Rev. Rel.}, vol.~11, p.~7, 2008.

\bibitem{Font07}
J.~A. Font, ``An introduction to relativistic hydrodynamics,'' {\em Journal of
  Physics: Conference Series}, vol.~91, p.~012002, nov 2007.

\bibitem{Carrasco17}
F.~L. Carrasco and O.~A. Reula, ``Novel scheme for simulating the force-free
  equations: Boundary conditions and the evolution of solutions towards
  stationarity,'' {\em Phys. Rev. D}, vol.~96, p.~063006, Sep 2017.

\bibitem{Carrasco18Pulsar}
F.~Carrasco, C.~Palenzuela, and O.~Reula, ``Pulsar magnetospheres in general
  relativity,'' {\em Phys. Rev. D}, vol.~98, p.~023010, Jul 2018.

\bibitem{Komiss04}
S.~S. Komissarov, ``{General relativistic magnetohydrodynamic simulations of
  monopole magnetospheres of black holes},'' {\em Monthly Notices of the Royal
  Astronomical Society}, vol.~350, pp.~1431--1436, 06 2004.

\bibitem{Fernandez18}
R.~Fernández, A.~Tchekhovskoy, E.~Quataert, F.~Foucart, and D.~Kasen,
  ``{Long-term GRMHD simulations of neutron star merger accretion discs:
  implications for electromagnetic counterparts},'' {\em Monthly Notices of the
  Royal Astronomical Society}, vol.~482, pp.~3373--3393, 10 2018.

\bibitem{Kovtun12}
P.~Kovtun, ``{Lectures on hydrodynamic fluctuations in relativistic
  theories},'' {\em J. Phys. A}, vol.~45, p.~473001, 2012.

\bibitem{Eckart40}
C.~Eckart, ``The thermodynamics of irreversible processes. iii. relativistic
  theory of the simple fluid,'' {\em Phys. Rev.}, vol.~58, pp.~919--924, Nov
  1940.

\bibitem{hiscock1983stability}
W.~A. Hiscock and L.~Lindblom, ``Stability and causality in dissipative
  relativistic fluids,'' {\em Annals of Physics}, vol.~151, no.~2,
  pp.~466--496, 1983.

\bibitem{Hiscock:1985zz}
W.~A. Hiscock and L.~Lindblom, ``{Generic instabilities in first-order
  dissipative relativistic fluid theories},'' {\em Phys. Rev.}, vol.~D31,
  pp.~725--733, 1985.

\bibitem{Israel-Stew79}
W.~Israel and J.~M. Stewart, ``{Transient relativistic thermodynamics and
  kinetic theory},'' {\em Annals Phys.}, vol.~118, pp.~341--372, 1979.

\bibitem{liu1986relativistic}
I.-S. Liu, I.~M{\"u}ller, and T.~Ruggeri, ``Relativistic thermodynamics of
  gases,'' {\em Annals of Physics}, vol.~169, no.~1, pp.~191--219, 1986.

\bibitem{geroch1991causal}
R.~Geroch and L.~Lindblom, ``Causal theories of dissipative relativistic
  fluids,'' {\em Annals of Physics}, vol.~207, no.~2, pp.~394--416, 1991.

\bibitem{Kreiss70}
K.~Heinz‐Otto, ``Initial boundary value problems for hyperbolic systems,''
  {\em Communications on Pure and Applied Mathematics}, vol.~23, no.~3,
  pp.~277--298, 1970.

\bibitem{geroch1996partial}
R.~Geroch, ``Partial differential equations of physics,'' {\em General
  Relativity, Aberdeen, Scotland}, pp.~19--60, 1996.

\bibitem{Hadamard1908}
J.~Hadamard, ``Théorie des équations aux dérivées partielles linéaires
  hyperboliques et du problème de cauchy,'' {\em Acta Math.}, vol.~31,
  pp.~333--380, 1908.

\bibitem{Friedrichs54}
F.~K. O., ``Symmetric hyperbolic linear differential equations,'' {\em
  Communications on Pure and Applied Mathematics}, vol.~7, no.~2, pp.~345--392,
  1954.

\bibitem{friedrichs1971systems}
K.~O. Friedrichs and P.~D. Lax, ``Systems of conservation equations with a
  convex extension,'' {\em Proceedings of the National Academy of Sciences},
  vol.~68, no.~8, pp.~1686--1688, 1971.

\bibitem{Aloy99}
M.~A. Aloy, J.~M. Ibanez, J.~M. Marti, and E.~Muller, ``{GENESIS}: A
  high-resolution code for three-dimensional relativistic hydrodynamics,'' {\em
  The Astrophysical Journal Supplement Series}, vol.~122, pp.~151--166, may
  1999.

\bibitem{Font00}
J.~A. Font, ``{Numerical hydrodynamics in general relativity},'' {\em Living
  Rev. Rel.}, vol.~3, p.~2, 2000.

\bibitem{Font02}
J.~A. Font, T.~Goodale, S.~Iyer, M.~Miller, L.~Rezzolla, E.~Seidel,
  N.~Stergioulas, W.-M. Suen, and M.~Tobias, ``Three-dimensional numerical
  general relativistic hydrodynamics. ii. long-term dynamics of single
  relativistic stars,'' {\em Phys. Rev. D}, vol.~65, p.~084024, Apr 2002.

\bibitem{Radice11}
D.~Radice and L.~Rezzolla, ``Discontinuous galerkin methods for
  general-relativistic hydrodynamics: Formulation and application to
  spherically symmetric spacetimes,'' {\em Phys. Rev. D}, vol.~84, p.~024010,
  Jul 2011.

\bibitem{Radice12}
{Radice, D.} and {Rezzolla, L.}, ``Thc: a new high-order finite-difference
  high-resolution shock-capturing code for special-relativistic
  hydrodynamics,'' {\em A\&A}, vol.~547, p.~A26, 2012.

\bibitem{Radice14}
D.~Radice, L.~Rezzolla, and F.~Galeazzi, ``High-order fully
  general-relativistic hydrodynamics: new approaches and tests,'' {\em
  Classical and Quantum Gravity}, vol.~31, p.~075012, mar 2014.

\bibitem{Lax73}
P.~Lax, {\em Hyperbolic Systems of Conservation Laws and the Mathematical
  Theory of Shock Waves}.
\newblock No.~n.{\textordmasculine} 11-16 in CBMS-NSF Regional Conference
  Series in Applied Mathematics, Society for Industrial and Applied
  Mathematics, 1973.

\bibitem{Roe97}
P.~Roe, ``Approximate riemann solvers, parameter vectors, and difference
  schemes,'' {\em Journal of Computational Physics}, vol.~135, no.~2, pp.~250
  -- 258, 1997.

\bibitem{Alcubierre08}
M.~Alcubierre, ``{Introduction to 3+1 Numerical Relativity},'' {\em Series:
  International Series of Monographs on Physics}, p.~ISBN: 9780199205677, April
  2008.

\bibitem{leveque92}
R.~J. LeVeque, {\em Numerical methods for conservation laws}, vol.~214.
\newblock Springer, 1992.

\bibitem{Harten83}
A.~Harten, ``High resolution schemes for hyperbolic conservation laws,'' {\em
  Journal of Computational Physics}, vol.~49, no.~3, pp.~357 -- 393, 1983.

\bibitem{Blazek01}
J.~Blazek, {\em Computational Fluid Dynamics: Principles and Applications}.
\newblock Elsevier Science, 2001.

\bibitem{Versteeg07}
H.~Versteeg and W.~Malalasekera, {\em An Introduction to Computational Fluid
  Dynamics: The Finite Volume Method}.
\newblock Pearson Education Limited, 2007.

\bibitem{LaxWen60}
P.~Lax and B.~Wendroff, ``Systems of conservation laws,'' {\em Communications
  on Pure and Applied Mathematics}, vol.~13, no.~2, pp.~217--237, 1960.

\bibitem{Thomas13}
J.~Thomas, {\em Numerical Partial Differential Equations: Finite Difference
  Methods}.
\newblock Texts in Applied Mathematics, Springer New York, 2013.

\bibitem{Lax-num1-54}
P.~D. Lax, ``Weak solutions of nonlinear hyperbolic equations and their
  numerical computation,'' {\em Communications on Pure and Applied
  Mathematics}, vol.~7, no.~1, pp.~159--193, 1954.

\bibitem{Bemfica21}
F.~S. Bemfica, M.~M. Disconzi, and P.~J. Graber, ``Local well-posedness in
  sobolev spaces for first-order barotropic causal relativistic viscous
  hydrodynamics,'' {\em Communications on Pure and Applied Analysis}, vol.~20,
  no.~9, pp.~2885--2914, 2021.

\bibitem{Bemfica:2017wps}
F.~S. Bemfica, M.~M. Disconzi, and J.~Noronha, ``{Causality and existence of
  solutions of relativistic viscous fluid dynamics with gravity},'' 2017.

\bibitem{Leray-Ohya67}
J.~Leray, ``{\'E}quations et syst{\`e}mes non-lin{\'e}aires, hyperboliques
  non-stricts,'' {\em S{\'e}minaire Jean Leray}, vol.~1964, no.~2, pp.~16--76,
  1965.

\bibitem{Leray53}
J.~Leray and N.~Institute~for Advanced Study~(Princeton, {\em Hyperbolic
  differential equations}.
\newblock Princeton Institute for Advanced Study, 1953.

\bibitem{friedrichs1954symmetric}
K.~O. Friedrichs, ``Symmetric hyperbolic linear differential equations,'' {\em
  Communications on pure and applied Mathematics}, vol.~7, no.~2, pp.~345--392,
  1954.

\bibitem{Pretorius21}
A.~Pandya and F.~Pretorius, ``Numerical exploration of first-order relativistic
  hydrodynamics,'' {\em Phys. Rev. D}, vol.~104, p.~023015, Jul 2021.

\bibitem{Pretorius22}
A.~Pandya, E.~R. Most, and F.~Pretorius, ``Conservative finite volume scheme
  for first-order viscous relativistic hydrodynamics,'' {\em Phys. Rev. D},
  vol.~105, p.~123001, Jun 2022.

\bibitem{Figueras22}
H.~Bantilan, Y.~Bea, and P.~Figueras, ``{Evolutions in first-order viscous
  hydrodynamics},'' {\em JHEP}, vol.~08, p.~298, 2022.

\bibitem{pennisi87}
S.~Pennisi, ``Some considerations on a non linear approach to extended
  thermodynamics,'' {\em Symposium of Kinetic Theory and Extended
  Thermodynamics}, Bologna, 1987.

\bibitem{geroch1995relativistic}
R.~Geroch, ``Relativistic theories of dissipative fluids,'' {\em Journal of
  Mathematical Physics}, vol.~36, no.~8, pp.~4226--4241, 1995.

\bibitem{geroch2001hyperbolic}
R.~Geroch, ``On hyperbolic "theories" of relativistic dissipative fluids,''
  {\em arXiv preprint gr-qc/0103112}, 2001.

\bibitem{Kremer02}
C.~Cercignani and G.~M. Kremer, {\em The Relativistic Boltzmann Equation:
  Theory and Applications}.
\newblock No.~PMP, volume 22 in Progress in Mathematical Physics book series,
  Birkh\"auser, Basel, 2002.

\bibitem{Kreiss1997GlobalEA}
H.-O. Kreiss, G.~Nagy, O.~E. Ortiz, O.~A. Reula, Ucla, U.~FaMAF, U.~N.
  de~Cordoba, and Argentina., ``Global existence and exponential decay for
  hyperbolic dissipative relativistic fluid theories,'' {\em Journal of
  Mathematical Physics}, vol.~38, pp.~5272--5279, 1997.

\bibitem{gowda2021high}
S.~Gowda, Y.~Ma, A.~Cheli, M.~Gwozdz, V.~B. Shah, A.~Edelman, and
  C.~Rackauckas, ``High-performance symbolic-numerics via multiple dispatch,''
  {\em arXiv preprint arXiv:2105.03949}, 2021.

\bibitem{doi:10.1137/070679065}
C.-W. Shu, ``High order weighted essentially nonoscillatory schemes for
  convection dominated problems,'' {\em SIAM Review}, vol.~51, no.~1,
  pp.~82--126, 2009.

\bibitem{SHU1988439}
C.-W. Shu and S.~Osher, ``Efficient implementation of essentially
  non-oscillatory shock-capturing schemes,'' {\em Journal of Computational
  Physics}, vol.~77, no.~2, pp.~439--471, 1988.

\bibitem{Palenzuela18}
C.~Palenzuela, B.~Mi\~nano, D.~Vigan\`o, A.~Arbona, C.~Bona-Casas, A.~Rigo,
  M.~Bezares, C.~Bona, and J.~Mass\'o, ``{A Simflowny-based finite-difference
  code for high-performance computing in numerical relativity},'' {\em Class.
  Quant. Grav.}, vol.~35, no.~18, p.~185007, 2018.

\bibitem{Bernuzzi16}
S.~Bernuzzi and T.~Dietrich, ``{Gravitational waveforms from binary neutron
  star mergers with high-order weighted-essentially-nonoscillatory schemes in
  numerical relativity},'' {\em Phys. Rev. D}, vol.~94, no.~6, p.~064062, 2016.

\bibitem{Liebling2020}
S.~L. Liebling, C.~Palenzuela, and L.~Lehner, ``{Toward fidelity and
  scalability in non-vacuum mergers},'' {\em Class. Quant. Grav.}, vol.~37,
  no.~13, p.~135006, 2020.

\bibitem{BORGES20083191}
R.~Borges, M.~Carmona, B.~Costa, and W.~S. Don, ``An improved weighted
  essentially non-oscillatory scheme for hyperbolic conservation laws,'' {\em
  Journal of Computational Physics}, vol.~227, no.~6, pp.~3191--3211, 2008.

\end{thebibliography}
	
\end{document}